\documentclass[pdflatex,sn-nature]{sn-jnl}

\usepackage{graphicx}%
\usepackage{multirow}%
\usepackage{amsmath,amssymb,amsfonts}%
\usepackage{amsthm}%
\usepackage{mathrsfs}%
\usepackage[title]{appendix}%
\usepackage[dvipsnames]{xcolor}%
\usepackage{textcomp}%
\usepackage{manyfoot}%
\usepackage{booktabs}%
\usepackage{algorithm}%
\usepackage{algorithmicx}%
\usepackage{algpseudocode}%
\usepackage{listings}%
\usepackage{comment}
\usepackage{enumerate}
\usepackage{lineno}
\usepackage{caption}

\newcommand{\software}[1]{{\textsc{#1}}}

\theoremstyle{thmstyleone}%
\theoremstyle{thmstyletwo}%

\theoremstyle{thmstylethree}%

\begin{document}

\title[TBD]{A possible challenge for Cold and Warm Dark Matter}


\author[1]{\fnm{Simona} \sur{Vegetti}}

\author[1]{\fnm{Simon D.~M.} \sur{White}}

\author[2,3,4]{\fnm{John P.} \sur{McKean}}

\author[1]{\fnm{Devon M.} \sur{Powell}}

\author[5]{\fnm{Cristiana} \sur{Spingola}}

\author[6]{\fnm{Davide} \sur{Massari}}

\author[6,7,8]{\fnm{Giulia} \sur{Despali}}

\author[9]{\fnm{Christopher D.} \sur{Fassnacht}}

\affil[1]{\orgname{Max Planck Institute for Astrophysics}, \orgaddress{\street{Karl-Schwarzschild-Stra\ss{}e 1}, \postcode{85748} \city{Garching bei M\"unchen},  \country{Germany}}}

\affil[2]{\orgdiv{Kapteyn Astronomical Institute}, \orgname{University of Groningen}, \orgaddress{\street{PO Box 800}, \postcode{NL-9700 AV} \city{Groningen}, \country{The Netherlands}}}

\affil[3]{\orgdiv{South African Radio Astronomy Observatory}, 
\orgaddress{\street{Krugersdorp 1740}, \postcode{P.O. Box 443},  \country{South Africa}}}

\affil[4]{\orgdiv{Department of Physics}, \orgname{University of Pretoria}, \orgaddress{\street{Lynnwood Road}, \city{Pretoria}, \postcode{0083},  \country{South Africa}}}

\affil[5]{\orgdiv{Istituto di Radioastronomia}, \orgname{INAF}, \orgaddress{\street{via Gobetti 101}, \postcode{I-40129} \city{Bologna}, \country{Italy}}}

\affil[6]{\orgdiv{Osservatorio di Astrofisica e Scienza dello Spazio}, \orgname{INAF}, \orgaddress{\street{via Gobetti 93/3}, \postcode{I-40129} \city{Bologna}, \country{Italy}}}

\affil[7]{\orgdiv{Dipartimento di Fisica e Astronomia}, \orgname{Alma Mater Studiorum Università di Bologna}, \orgaddress{\street{via Gobetti 93/2}, \postcode{I-40129} \city{Bologna}, \country{Italy}}}

\affil[8]{\orgdiv{Sezione di Bologna}, \orgname{INFN}, \orgaddress{\street{ Viale Berti Pichat}, \postcode{I-40127} \city{Bologna}, \country{Italy}}}

\affil[9]{\orgdiv{Department of Physics and Astronomy}, \orgname{University of California Davis}, \orgaddress{\street{1 Shields Avenue},  \city{Davis}, \state{CA},  \country{USA}, \postcode{95616}}}

\maketitle
\newpage
\noindent{{\bf Measuring the density profile and mass concentration of dark-matter haloes is a key test of the standard cold dark matter paradigm. Such objects are dark and thus challenging to characterise, but they can be studied via gravitational lensing. Recently, a million-solar-mass object was discovered superposed on an extended and extremely thin gravitational arc. Here we report on extensive tests of various assumptions for the mass density profile and redshift of this object. We find models that best describe the data have two components: an unresolved point-mass of radius $\leq10$ pc centred on an extended mass distribution with an almost constant surface density out to a truncation radius of 139~pc. These properties do not resemble any known astronomical object. However, if the object is dark matter-dominated, its structure is incompatible with cold dark matter models, but may be compatible with a self-interacting dark matter halo where the central region has collapsed to form a black hole. This detection could thus carry substantial implications for our current understanding of dark matter.}
\vspace{\baselineskip}

Dark matter makes up 85\% of cosmic matter, but its nature is still unknown. The Cold Dark Matter (CDM) paradigm, in which dark matter consists of massive non-relativistic, collisionless elementary particles, agrees well with a wide variety of astrophysical observations. However, it remains largely untested on sub-galactic scales. One prediction of this model is that cosmic structure formed through a hierarchical bottom-up process. As a result, a large population of low-mass dark-matter haloes (down to planetary mass scales) is expected to exist both (1) as sub-haloes in massive galaxy and cluster haloes and (2) in the field, with mass functions and mass density profiles that have been well characterised through extensive numerical simulations \citep[e.g.][]{springel_2008}. Modifications of CDM in which dark matter particles have a non-negligible thermal velocity at early times (i.e. warm dark matter; WDM) can predict orders of magnitude fewer low-mass haloes, and these have significantly less concentrated mass density profiles \citep[e.g.][]{lovell_2012, lovell_2020}. The reduction of the number of haloes and their concentration is directly related to the mass of the WDM particle.

Self-interacting dark matter (SIDM), where the particles have both gravitational and non-gravitational interactions, can redistribute energy and momentum within haloes, modifying their dark matter distribution and thus creating more diverse density profiles than in CDM \citep[e.g.,][]{kahlhoefer_2019, zavala_2019, nishikawa_2020, turner_2021, zeng_2022}. If the self-interaction cross-section is high enough, haloes can evolve to have a highly concentrated density profile at the centre. Such objects form via a runaway contraction process, known as the gravo-thermal catastrophe, or core-collapse \citep{lynden-bell_1968, balberg_2002,outmezguine_2023}, and can be considerably more centrally concentrated than CDM haloes of the same mass. Core collapse rapidly leads to the formation of a central black hole, which, unlike in CDM models, is not dependent on complex and uncertain baryonic processes. Therefore, directly measuring the number and density profile of dark matter haloes at low masses (below 10$^8 M_{\odot}$, where $M_\odot$ is the mass of the Sun) can robustly discriminate between dark matter models. These objects are expected to be dark matter dominated, and most should be completely dark. They can thus only be studied with a gravitational probe, such as strong gravitational lensing.

The JVAS B1938+666 system comprises a massive elliptical galaxy at redshift $z_{\rm lens} = 0.881$ that gravitationally lenses a powerful radio source at redshift $z_{\rm src} = 2.059$ \citep{king_1997,king_1998,riechers_2011,lagattuta_2012,spingola_2020}. At near-infrared wavelengths (2.1$~\mu$m; observed-frame), the host galaxy of the radio source forms an almost complete Einstein ring \citep{king_1998,lagattuta_2012}, against which, a $(1.9 \pm 0.1) \times 10^8~M_\odot$ dark object has been detected via its gravitational lensing effect \citep{vegetti_2012}. The radio source was observed with the global Very Long Baseline Interferometry (VLBI) network at 1.7 GHz (observed frame), revealing a spectacular and very thin gravitational arc extending over about 200 milliarcseconds with a width of at most a few milliarcseconds \citep{mckean_2025}. This arc is clearly separated on the sky from the infrared Einstein ring, and a gravitational imaging analysis of the 1.7 GHz visibility data by Powell et al. \citep{powell_2025} detected (at $26\sigma$) the gravitational effects of a second perturber (detection ${\cal V}$) that has no obvious luminous counterpart in near-infrared W. M. Keck Telescope adaptive optics data. \citet{powell_2025} estimated a mass of $\sim10^6~M_{\odot}$ for this object under the assumptions that it lies within the lens galaxy and that its mass profile is well described by a truncated singular isothermal sphere. For clarity, we note that \emph{gravitational imaging} refers to the lens modelling technique in which low-mass haloes are \emph{imaged} as regularised pixellated corrections to the otherwise smooth lensing potential. This concept was first introduced by \citet{koopmans_2005}. The term is sometimes erroneously used to describe detections obtained with methods that assume a parametrized model for the perturber mass distribution.

We apply the lens modelling code \software{pronto} \citep{vegetti_2009,rybak_2015,rizzo_2018, ritondale_2019,powell_2021,ndiritu_2024}, extended to allow the analysis of high-angular resolution interferometric data, to investigate the likely nature of detection ${\cal V}$. We test and compare a wide variety of different parametric models for its mass density profile and redshift using the logarithmic Bayesian evidence ($\ln{\cal E}$, that is, the natural logarithm of the probability of a model given the data). These models are chosen to span the most likely candidates for the nature of this unprecedentedly low-mass object: a compact nucleus or black hole, described by a point mass; a globular cluster, described by a King or Plummer profile; a dark matter halo or subhalo, described by a conventional Navarro–Frenk–White profile, a truncated singular isothermal profile or a broken power-law profile; and an ultra-compact dwarf galaxy, described by a Sérsic profile. We consider both uninformative priors for the parameters and priors based on observed systems. We also consider composite models that superpose an unresolved central component (a black hole or compact nucleus) on an extended component. We also compare models with the perturber within the lens (at a redshift of 0.881) to models where it is at an unknown redshift along the line of sight. 

In total, we tested 23 different models (Tables~\ref{tab:main_tab1} and \ref{tab:main_tab2} and Methods). The best-fitting model (Figure~\ref{fig:main_fig1} and Extended Data Table \ref{tab:ed_tab1}) is an object at the main lens redshift made up of a point mass and a uniform, face-on disc (or the equivalent limits of the Sérsic and broken power-law profiles). We designate this as our reference model and give the evidence for other models as $\Delta\ln {\cal E}$ relative to it. However, we note that this model has a sharp surface-density cut-off at the boundary (139 pc), which is challenging to reconcile with the profiles of known astrophysical systems. This may reflect some inadequacy of our main lens model at these small scales, or some un-modelled line-of-sight effect that modifies the surface brightness of the radio arc on scales larger than the low mass perturber. The best astrophysically plausible model we have found for the extended component is a Plummer sphere, but this is already disfavoured by $\Delta\ln {\cal E}=-17$. A model with no perturber has $\Delta\ln {\cal E}= -388$, corresponding to 28$\sigma$ for Gaussian statistics.

Adopting a single point mass model for the perturber leads to a total mass of $M_{\rm{t}} = (1.450\pm0.066)\times10^6~M_{\odot}$, but the data strongly disfavour this model with $\Delta\ln{\cal E}= -128$. Hence, our lensing analysis rules out a black hole or unresolved ``point'' mass interpretation at high statistical significance. The number density of globular clusters around the Einstein radius of elliptical galaxies is, in some cases, comparable to that of dark-matter sub-haloes (Supplementary Information). Depending on the adopted prior on cluster parameters, the VLBI data disfavour a King model with a $\Delta\ln{\cal E}$ between $-52$ and $-122$. The best-fitting King profile with a finite truncation radius has a mass of $\left(3.05\pm0.57\right)\times10^6M_\odot$, a core radius of $r_{\rm c}= 9.39\pm1.21$ pc and a tidal radius of $r_{\rm t} =24.7\pm22.4$ kpc.  These properties are extreme compared to known globular clusters, but not outside the observed range (Extended Data Figure \ref{fig:ed_fig1}). Given the values of the Bayesian evidence, we conclude that the detected object is very unlikely to be a globular cluster in the main lens.

On the other hand, the best-fitting single-component NFW profiles \citep{navarro_1996} both for a field halo (at $z=1.469$) and for a sub-halo in the lens are highly concentrated, with $\log c_{\rm vir} = \log (r_{\rm vir}/r_{\rm s})= 2.91\pm0.09$ and $\log c_{\rm v} = \log\left( 2\times\left({V_{\rm max}}/\left(H_z r_{\rm max}\right)\right)^2\right)=7.25\pm0.17$, respectively. For comparison, the expected concentrations at these masses and redshifts in a $\Lambda$CDM cosmology are $\log c_{\rm vir}=1.10\pm 0.15$ \citep{duffy_2008} and $\log c_{\rm v}=3.90\pm0.02$ \citep{moline_2023}. Hence, we find a discrepancy at the 10 to 20$\sigma$ level between our observations and theoretical expectations. Imposing such a $\Lambda$CDM prior leads to a dramatic drop in Bayesian evidence of $-75$ and $-147$ for the field halo and subhalo cases, respectively (see Extended Data Figure \ref{fig:ed_fig2} for a comparison of the measured and predicted maximum radius and velocity). Adding a central point mass reduces the tension between the inferred NFW parameters and the $\Lambda$CDM expectation to 3 and $6\sigma$, but this has little meaning since adding a point mass of the mass we infer would substantially modify the central profile of the host halo. Even with the NFW parameters and the central point mass left free, all NFW models for the extended component are strongly excluded by the data, with the best model of this family (for which the mass of the compact object is $0.3\times M_{\rm max}$) having $\Delta\ln{\cal E}=-31$ relative to our best-fit model.

In our analysis, detection ${\cal V}$ is best explained by a uniform-surface-density, face-on disk of outer radius $R= 139\pm4$ pc centred on an unresolved component containing 19\% of the total mass, $(1.8 \pm 0.1)\times 10^6M_\odot$.
The central object could potentially be a black hole or a nuclear star cluster. In CDM and WDM, the formation of a black hole at the centre of galaxies is the result of complex and uncertain baryonic processes. In the mass regime of detection $\cal{V}$, haloes are expected to have no stars and to be well described by a single NFW profile. A fully dark halo with properties consistent with our best-fit model thus seems extremely implausible in CDM and WDM models (see Supplementary Information for more details). However, for suitably chosen cross-sections, it may be achievable in SIDM models, where gravothermal evolution and core collapse can lead to the formation of black holes at the centre of dark matter haloes \citep[e.g.][]{turner_2021, outmezguine_2023, yang_2024}. Reproducing the VLBI data would require a halo that had a CDM equivalent with a maximum circular velocity $V_{\rm max}\sim 5$ km\,s$^{-1}$ and that is deep in the core-collapse regime by the observed redshift of 0.881, implying a velocity averaged interaction cross-section $\sim 800$ cm$^2$~g$^{-1}$, or more (Extended Data Figure  \ref{fig:ed_fig3} and Supplementaty Information).

Alternatively, the lensing signature of detection ${\cal V}$ could be that of an ultracompact galaxy with a central massive black hole or nuclear star cluster. These are a distinct class of small, dense stellar systems with masses lying between those of globular clusters and those of normal dwarf galaxies. First discovered in nearby galaxy clusters such as Fornax and Virgo \citep[e.g.][]{hilker_1999}, they are some of the densest galaxies currently known. Ultracompact galaxies are thought to form in a variety of ways, for example, by tidal stripping of dwarf galaxies in dense environments. Our best-fit model has an overall size and mass consistent with those of known ultracompact galaxies and local nuclear star clusters  (Extended Data Figure  \ref{fig:ed_fig1} and Supplementary Information), although their radial light distributions are markedly more centrally concentrated than the uniform mass surface density favoured here.

This is the third object to be individually \emph{gravitationally imaged} to date \citep{vegetti_2010,vegetti_2012}. All three detections have properties which, at different statistical levels, appear unusual when compared to dark-matter-dominated haloes in the CDM and WDM paradigms  \citep{minor_2021, despali_2024, enzi_2024, tajalli_2025}. Our analysis (see Supplementary Information for more details), together with those of previous detections, suggests that if deeper observations confirm that these objects are dark-matter-dominated rather than star-dominated, as ultracompact galaxies and nuclear star clusters are, then dark matter cannot be collisionless. Hence, our results could have important implications for the nature of dark matter and the standard cosmological model. However, more numerical and theoretical work is needed to obtain robust predictions from SIDM models at the spatial scales and evolutionary phases relevant to the VLBI data, and to verify that all three detections are consistently explained by the same parameters for the SIDM cross-section.

\section*{Methods}
\label{method}

\subsection*{Data}

The data for the radio source JVAS~B1938+666 used in this analysis were taken with a global VLBI array consisting of the European VLBI Network (EVN, 11 antennas), the Very Long Baseline Array (VLBA, 10 antennas) and the Green Bank Telescope, at a central observing frequency of 1.7 GHz and with a bandwidth of 64 MHz \citep{mckean_2025}. This observational set-up resulted in an angular resolution (the so-called beam size) of 7.4 mas $\times$~4.7~mas, for a natural weighting of the visibilities. The observation lasted 14 h with a data recording rate of 512 Mbits s$^{-1}$ (ID: GM068, principal investigator McKean).  

Using these data, an extended, but extremely thin gravitational arc was found \citep{mckean_2025}, superposed on which a low-mass object was discovered by Powell et al. \citep{powell_2025}. Here, we focus on the properties of this new object (detection ${\cal V}$) and consider possible explanations for its nature in the context of known astrophysical objects and various dark-matter models.

\subsection*{Lens model}

We use the Bayesian lens modelling code \software{pronto} \citep{vegetti_2009, rybak_2015, rizzo_2018, ritondale_2019, powell_2021, ndiritu_2024}, which fits the data directly in visibility space. The background unlensed source is reconstructed in a pixellated, regularised fashion on a Delaunay tessellation with a magnification-adaptive resolution. The sum of the following components gives the mass density profile of the macro-model: an elliptical power law for the main lens galaxy, multipoles of order $m = 3$ and $m=4$ (as defined by \cite{oriordan_2024}), and an external shear component of strength $\Gamma$ and position angle $\Gamma_{\theta}$. We also include the contribution of a $\sim 10^8 M_\odot$ low-mass halo, first detected by \cite{vegetti_2012} in the same gravitational lens system (in the following referred to as detection ${\cal A}$). \citet{powell_2025} have shown that, while the VLBI data do not provide a strong constraint on the redshift of detection ${\cal A}$, they strongly disfavour the $z=1.4$ value proposed by \cite{sengul_2022}. Recently, \citet{tajalli_2025} have presented a re-analysis of the W. M. Keck Telescope adaptive optics (Keck AO hereafter) data for JVAS B1938+666, showing that detection ${\cal A}$ is indeed a field halo, but at much lower redshift, $z=0.13\pm 0.07$. We note that the Keck AO data, when compared with the VLBI data analysed here, are significantly more sensitive to the presence and properties of detection ${\cal A}$, as it falls right on top of the lensed (rest-frame) optical emission, while offset relative to the 1.7-GHz arc. In fact, it was demonstrated in ref. \citep{powell_2025} that assumptions about detection ${\cal A}$ do not affect the detection and the inferred properties of the object under study here (which we refer to as detection ${\cal V}$). Hence, for simplicity, we include detection ${\cal A}$ as a pseudo-Jaffe sub-halo at the lens redshift. 

The dimensionless surface mass density (that is, the convergence) profile for an elliptical power law of axis ratio $q$ and three-dimensional (3D) slope $\gamma$  (with $\gamma =2$ representing an isothermal profile) is given by
\begin{equation}
\kappa(R) = \kappa_0\left(1-\frac{\gamma}{4}\right)\frac{q^{\gamma-3/2}}{(R/r_{\rm p})^{(\gamma-1)}}\,,
\label{eq:pl_kappa}
\end{equation}
with radius $R=\sqrt{q^2(r_{\rm c}^2+x^2)+y^2}$ and normalisation $\kappa_0$. We calculate the corresponding deflection angle using \software{fastell} \citep{barkana_1999}. With the exception of the core radius ($r_{\rm {c}}\equiv 10^{-4}$ arcsec), the pivot radius ($r_{\rm p}\equiv 1$ arcsec) and the redshift of detection ${\cal A}$ ($z_{{\cal A}}\equiv z_{\rm lens}$) all other parameters of the macro-model are left free to vary (for a total of 16 free parameters including the strength of the source regularisation). 

We model detection ${\cal V}$ with the following circularly symmetric profiles: a single power law (PL), a broken power law (bPL), a truncated singular isothermal or pseudo-Jaffe (PJ) profile, a Navarro-Frenk-White (NFW) profile, a King model (KG), a Sérsic (SER) profile, a Plummer (PLU) profile and a point mass (PM). We also tested an elliptical power law (ePL) profile and a number of composite models that add a central point mass to one of the above models for the extended component. In all cases, we keep the position on the sky free.

The convergence of the PL model follows equation (\ref{eq:pl_kappa}) with $q=1$. Free parameters of the model are the normalisation $\kappa_0$ and the slope $\gamma$. For the core radius, we choose a very small value that is well below the resolution of the data. We consider both a sub-halo model (PL) where the redshift is fixed to that of the main deflector and a field halo model (PL$_{\rm field}$) where the redshift is a free parameter.

For the broken power law model, we adopt the representation of ref. \cite{Riordan_2021} for the convergence
 \begin{equation}
\kappa(x) = 
\begin{cases}\displaystyle\kappa_{\rm b}~x^{-t_1}   & \mathrm{for}\,\, x \le 1\\
\displaystyle\kappa_{\rm b}~x^{-t_2}               &\mathrm{for}\,\,  x > 1\\
\end{cases}
\,,
\label{eq:bpl-convergence}
\end{equation}
where $x = R/r_{\rm b}$. The break radius $r_{\rm b}$, the inner and outer logarithmic slopes $t_1$ and $t_2$, and the normalisation at the break radius $\kappa_{\rm b}$ are all adjustable parameters.

The PJ profile \citep{jaffe_1983} is described as follows: 
\begin{equation}
\kappa(x) = \kappa_0\left(\frac{1}{x}-\frac{1}{\sqrt{1+x^2}}\right)\,,
\label{eq:pj_density}
\end{equation}
with $x=R/r_{\rm t}$. This model has two degrees of freedom, that is, the total mass $M_{\rm t}=2 \pi r_{\rm t}^2\kappa_0\Sigma{\rm _c}$ and the truncation radius $r_{\rm t}$, where $\Sigma{\rm _c}$ is the critical surface-mass density.

The convergence of the NFW \citep{navarro_1996} profile is given by
\begin{equation}
\kappa(x) = 2\kappa_{\rm s}\frac{1-F(x)}{x^2-1}\,, 
\label{eq:nfw_density}
\end{equation}
with $x=R/r_{\rm s}$,  $\kappa_{\rm s} = \rho_{\rm s} r_{\rm s}\Sigma_{\rm c}^{-1}$ and
\begin{equation}
F(x) = \begin{cases}
    {1 \over \sqrt{x^2-1}}\,\mbox{tan}^{-1} \sqrt{ x^2-1}  &  \mathrm{for}\,\,   x>1 \cr
    {1 \over \sqrt{1-x^2}}\,\mbox{tanh}^{-1}\sqrt{ 1-x^2 } &  \mathrm{for}\,\,   x<1 \cr
    1                                                      &  \mathrm{for}\,\,    x=1\cr 
  \end{cases}
  \,.
\label{eq:nfw_fx}
\end{equation}
Here, the density $\rho_s$ and the surface-mass density $\kappa_s$ are taken at the scale radius $r_s$. We consider two cases: one in which the object is a sub-halo within the lens galaxy (NFW) and one in which its redshift can vary between the observer and the source (NFW$_{\rm field}$). For sub-haloes, we characterise the profile by the peak circular velocity radius $r_{\rm max}$ and $M_{\rm max}$, the corresponding enclosed mass in 3D; further, we compare to N-body simulations using the concentration $c_{\rm v} = 2\times\left(V_{\rm max}/\left(H_z r_{\rm max}\right)\right)^2$, where $V_{\rm max}$ is the peak circular velocity and $H_z$ is the Hubble constant at redshift $z$. For field haloes, we use the more conventional parameter pair: the virial mass $M_{\rm vir}$ and the concentration $c_{\rm vir}=r_{\rm vir}/r_{\rm s}$. For both the PL$_{\rm field}$ and the NFW$_{\rm field}$ models, the prior on the redshift is set to be uniform in the comoving distance of the perturber. 

The King model is used to represent globular cluster (GC) profiles. Its convergence is given by the following expression
\begin{equation}
\kappa(x) = \kappa_0\left[\frac{1}{\sqrt{1+x^2}} - \frac{1}{\sqrt{1+t^2}}\right]\,, 
\label{eq:kg_kappa}
\end{equation}
where $x=R/r_{\rm c}$ and $t=r_{\rm t }/r_{\rm c}$. The degrees of freedom in this case are the core radius $r_{\rm c}$, the truncation (or ``tidal") radius $r_{\rm t}$ and the total mass. The latter is given by $M_{\rm t}=\pi\kappa_0 r_{\rm c}^2\Sigma_{\rm c}f(t)$, with
\begin{equation}
f(t) = \log(1+t^2)+\frac{t^2 - 4 \sqrt{1+t^2}\left(\sqrt{1+t^2}-1\right)}{1+t^2}\,.
\label{eq:kg_intg}
\end{equation}
We consider three cases. In the first, the truncation radius $r_{\rm t}$ is taken to be equivalent to the tidal radius, estimated using the expression \citep{bellazzini_2004}
\begin{equation}
r_t = \frac{2}{3}\left(\frac{M_{\rm {t}}}{ 2 M_{\rm {lens}}(<d)} \right)^{1/3} d\,.
\end{equation}
Here, $d$ is the projected distance of the detected object from the centre of the lens galaxy and $M_{\rm {lens}}(<d)$ is the mass of the main deflector within a sphere of radius $d$. We refer to this model as KG$_{\rm td}$ and set the prior on the concentration $c =\log{r_{\rm t}/r_{\rm c}}$ to be uniform between 0.3 and 2.7, in line with typical values for GCs in the Local Universe \citep{harris_2010}. For the second case (KG), we leave the total mass, the core radius, and $t^{-1}=r_{\rm c}/r_{\rm t}$ free to vary, with the prior on the latter being uniform between $10^{-4}$ and 0.5 to encompass all known values for GCs and ultra-compact dwarf galaxies (UCDs; see below for more details). For the third version of the King profile (KG$_{\infty}$) we allow $r_{\rm t}\rightarrow\infty$. As the total mass of this profile is not finite, we define as free parameters the mass $M_{\rm max}$ within the peak circular velocity radius  $r_{\rm max}\sim 2.919\times r_{\rm c}$ and the core radius itself, $r_{\rm c}$. 

The majority of UCDs have been shown to be well described by a multi-component King or Sérsic profiles \citep[see, for example, ref.][]{wang_2023}. The convergence profile of the latter is given by
\begin{equation}
\kappa(x) = \kappa_{\rm e}\exp{\left[-b_{\rm n}(x^{1/n}-1)\right]}\,,
\label{eq:kappa_ser}
\end{equation}
with $x=R/R_{\rm e}$. There are three degrees of freedom: the effective radius $R_{\rm e}$, the index $n$ and the total mass $M_{\rm t} = 2\pi R_{\rm e}^2 \kappa_{\rm e}\Sigma_c n \exp{\left(b_{\rm n}\right)}/b_{\rm n}^{2n}\Gamma\left[2 n\right]$. Here, $\Gamma\left[2 n\right]$ is the complete gamma function. The constant $b_n$ is set by the requirement that half the total mass be projected within $R_e$. In the limit $n\rightarrow 0$
\begin{equation}
\kappa (x) = \begin{cases}
    \frac{M_{\rm t}}{2} x^2 &   \mathrm{for}\,\, x < \sqrt{2} \cr
    M_{\rm t}               &   \mathrm{for}\,\, x \geq  \sqrt{2} \cr
\end{cases}
 \,.
\end{equation}
Hence, in this case, the Sérsic profile has a constant surface mass density and a finite size, and we refer to it as a uniform disk (UD). This model has only two degrees of freedom: the total mass and the effective radius. It is also a limiting case of the bPL model for $t_1=0$ and  $t_2\rightarrow\infty$.

The Plummer profile (PLU) was initially introduced to describe the light distribution of globular clusters \citep{Plummer_1911} and is now often used as an example of a simple, fully analytic model for spherical stellar systems. We consider it because, with an additional central point mass, it turns out to be a relatively good description of our data. The convergence of the Plummer model is given by
\begin{equation}
\kappa(x) = \frac{\kappa_{\rm 0}}{\left(1+x^2\right)^2}\,,
\label{eq:kappa_plu}
\end{equation}
with $x=r/r_{\rm s}$. We take the free parameters of this model to be the total mass $M_{\rm t}$ and the scale radius $r_{\rm s}$.

\subsection*{Model ranking}
\label{results}

Tables \ref{tab:main_tab1} and \ref{tab:main_tab2} list the mean and 1$\sigma$ uncertainty on the inferred parameters for all the different profiles that we tested. When the perturbers are located at a higher redshift than the main lens, we quote their position on the latter plane. Regardless of the assumed mass density profile, the object is detected at high statistical significance, with a Bayes factor relative to a model with no perturber of $241 \leq \Delta\ln{{\cal E}_s} \leq 388$, corresponding to a detection significance between 22 and 28$\sigma$ assuming Gaussian statistics. 

The best-fitting model is the UD+PM model. In the following, the Bayes factor of a given model $\cal{M}$ is measured, unless otherwise specified, relative to this best model, that is,  $\Delta\ln{\cal{E}}=\ln{\cal{E}_{\cal M}}-\ln{\cal{E}_{\rm UP + PM}}$.
Extended Data Table  \ref{tab:ed_tab1} reports the inferred parameters for the macro-model (including detection ${\cal A}$). We only quote these parameters for the best model (UD +PM), but we find that they change very little for different models for the profile of detection ${\cal V}$. Images of the system at a variety of scales, showing the position of detections ${\cal A}$ and ${\cal V}$ relative to the lensed source emission in the radio and the near-IR can be found in fig. 1 of \citet{powell_2025}. 

For models that do not include the PM component, we find small differences in the Bayes factor (at most 4) between those with the perturber at the redshift of the lens and those where its redshift is free. The latter converge to $z\sim 1.4$ and are characterised by steeper slopes for the PL models and higher concentrations for the NFW profiles. This is related to the degeneracy between these quantities and redshift, whereby objects located along the line-of-sight need to be more massive and to have steeper or more concentrated profiles to provide the most similar lensing effect to an object within the lens (Tajalli et al. in prep.). Once a PM component is included, the field models converge to a redshift consistent with that of the main lens, and their Bayes factor increases considerably. This indicates that detection $\cal{V}$ is very likely within the main deflector. In the following, we therefore focus primarily on sub-halo models with $z\equiv z_{\rm lens}$.

The top model UD+PM is followed closely by SER+PM ($\Delta\ln{{\cal E}}=-3.5$) and bPL+PM ($\Delta\ln{{\cal E}}=-11$). This is not surprising, since the parameters found for these two models are close to the limits where their mass profiles become identical to that of UD+PM. However, all three models are bounded by an implausibly steep surface mass density cut-off. The fourth best model is PLU+PM ($\Delta\ln{{\cal E}}=-17$). This result is also unsurprising since equation (\ref{eq:kappa_plu}) shows the convergence of this profile to be almost constant out to the scale radius, where it cuts off steeply. Sub-halo models with a point mass and an extended profile more similar to those usually assumed for dark-matter haloes or real stellar systems are significantly disfavored relative to these top models, with Bayes factors ranging from $\Delta\ln{{\cal E}}=-30$ for PJ+PM to $\Delta\ln{{\cal E}}=-40$ for ePL+PM. The best model without a central point mass is bPL with $\Delta\ln{{\cal E}}=-27$, followed by PL, SER, PJ and NFW in that order, with the last having a high concentration and $\Delta\ln{{\cal E}}=-44$; when a $\Lambda$CDM concentration prior is adopted, this drops catastrophically to $\Delta\ln{{\cal E}}=-147$. All versions of the King profile are strongly disfavoured, with Bayes factors ranging from $-52$ for KG$_{\infty}$ to $-122$ for KG$_{\rm td}$. A point mass is also very strongly excluded with $\Delta\ln{{\cal E}}=-128$. Hence, from the lens modelling alone, we conclude that the data exclude a naked black hole or a compact system with a radius less than 10 or 20~pc. In the Supplementary Information,  we discuss how these profile properties compare to predictions from different dark-matter models and to the observed properties of GCs and UCDs. 

Strong gravitational lensing is known to provide a robust and model-independent measure of the projected mass within an appropriately chosen radius. Our results show, however, that in the system we consider here, the projected mass profile is tightly constrained over a broad range of radii. This is a consequence of the fact that, for low-mass perturbers, the angular size of the source is larger than that of the deflector, so that the observed image structure constrains the enclosed mass over a range of radii. In the specific case of detection $\cal{V}$, we find all models that provide a decent fit to the data agree in the values they predict for the enclosed (cylindrical) mass, $M_{\rm cyl}$, both at 20 and (most tightly) at 90 pc. This applies to perturbers both within and along the line of sight to the main lens. Indeed, of the 12 models with  $\Delta\log{\cal E}>-40$, 10 give a best value of $M_{\rm cyl,90}$ that differs by less than 3\% from that of the best model, while for $M_{\rm cyl,20}$, 10 models differ from the best model by less than 8\%. 

Models such as KG$_{\rm td}$ and PM that do not lead to the correct mass at either radius, or the NFW$_{\rm CDM}$ model, which fails to match at 20 pc, are strongly disfavoured. Leaving aside these models, the ranking of the other profiles depends both on the curvature of the  $M_{\rm cyl}$ profile between 20 and 90 pc and on the rapidity with which the mass increases at larger radii. Thus, the gradient of the UD+PM and PLU+PM models is increasing outwards, while for the models with $-27>\Delta\log{\cal E}> -32$, it is constant over this range, and for most of the less favoured models it is decreasing outwards (Fig. \ref{fig:main_fig1} and Supplementary Fig. \ref{fig:si_fig1}). We demonstrate explicitly that the data also constrain the profile beyond 90 pc by running a bPL+PM model for which the inner point mass, the break radius and the outer slope were left free, but the inner slope was fixed to that of the best PL+PM model. We found this model to predict an essentially identical $M_{\rm cyl}(R)$ to PL+PM within the break radius, 98 pc, but a markedly steeper profile at larger radii ($t_2 = 1.96$ as compared to $t_1 = 1.155$). The $\Delta\log{\cal E}$ for this extra model was $-32$, a full 7 units better than the PL+PM model, indicating that a reduced mass beyond 100 pc is strongly preferred by the data. Presumably, this reflects the fact that the lensed arc extends well beyond this distance, and its surface brightness distribution provides extra constraining power.

Irrespective of the choice of model, the position of the perturber on the plane of the lens is extremely well-constrained, with an average error of $\sigma_x=0.1$ mas and  $\sigma_y=0.2$ mas (equivalent to 0.8-1.6 pc in the plane of the lens). The error on the $x$ coordinate is smaller by a factor of 2 because of the corresponding smaller size of the lensed arc in the $x$ direction, a result of the strong tangential magnification experienced by the background source, together with the location of the critical curves; any small shift of the perturber along $x$ corresponds to a substantial change in its gravitational effect. The position is also consistent among the different models, with a standard deviation of 1.1 and 1.5 mas along the $x$ and $y$ axes, respectively. The largest shift relative to our best model is displayed by the PM model, with $|\Delta x| = 5$ mas and $|\Delta y| = 4$ mas. This is related to the fact that the PM is the most centrally concentrated profile we consider, and so, has the largest capability to perturb the lensed images. If located closer to the arc, a PM would further split the arc into further multiple images, but there is no evidence for this in the observational data. 

Supplementary Fig. \ref{fig:si_fig1} compares the $M_{\rm cyl}(R)$ curves for all our profile models, while Supplementary Figs. \ref{fig:si_fig2}-\ref{fig:si_fig24} display the full posterior distributions for their parameters. 

\newpage

\backmatter

\section*{Declarations}

\bmhead{Acknowledgments} 

 S.~V. and D.~M. are grateful to Laura V. Sales, Eric Peng and Francesco Calura for kindly sharing the data required to reproduce Fig. 2 of \citet{wang_2023} and Fig. 12 of \citet{calura_2022}, respectively. S.~V. thanks Alessandro Zocco for providing clever computational solutions. S.~V. thanks Haibo-Yu, Daneng Yang and Shin'ichiro Ando for insightful discussion on SIDM. This research was carried out on the High-Performance Computing resources of the Raven cluster at the Max Planck Computing and Data Facility (MPCDF) in Garching, operated by the Max Planck Society (MPG).  S.~V., and D.~M.~P. have received funding from the European Research Council (ERC) under the European Union's Horizon 2020 research and innovation programme (grant agreement No 758853). S.~V. thanks the Max Planck Society for support through a Max Planck Lise Meitner Group. This work is based on the research supported in part by the National Research Foundation of South Africa (Grant Number: 128943). C.~S. acknowledges financial support from INAF under the project “Collaborative research on VLBI as an ultimate test to $\Lambda$CDM model” as part of the Ricerca Fondamentale 2022. This work was supported in part by the Italian Ministry of Foreign Affairs and International Cooperation, grant number  PGRZA23GR03 and by the Italian Ministry of University and Research (grant number FIS-2023-01611 - CUP C53C25000300001). D.~M. acknowledges financial support from PRIN-MIUR-22: CHRONOS: adjusting the clock(s) to unveil the CHRONO-chemo-dynamical Structure of the Galaxy” (PI: S. Cassisi).  G. D. acknowledges the funding by the European Union - NextGenerationEU, in the framework of the HPC project – “National Centre for HPC, Big Data and Quantum Computing” (PNRR - M4C2 - I1.4 - CN00000013 – CUP J33C22001170001).

\bmhead{Author contributions statement} 

S.~V. has contributed to the development of the lens modelling code and has carried out all the analysis presented in the paper. S.~D.~M.~W. has derived the fit to the SIDM profiles and contributed extensively to the interpretation of the results. J.~P.~M. initiated and led the data collection, and contributed to the calibration and imaging. D.~M.~P. was the main developer of \textsc{pronto} and has contributed to the smooth modelling of the data. C.~S. has contributed to this work by calibrating the VLBI data and writing the Data section. D.~M. has contributed to the interpretation of the results in terms of globular cluster properties. G.~D. has contributed to the interpretation of the results and discussions on SIDM predictions. C.~D.~F. has contributed to the interpretation of the results and obtained the Keck AO data that were used for the first detection of ${\cal A}$.

\bmhead{Data availability}

 The dataset is publicly available on the EVN archive https://archive.jive.nl/scripts/portal.php (bbservation ID: GM068, principal investigator McKean).

\bmhead{Code availability} 

The modelling code \software{pronto} is not publicly available. The reader interested in using this code can contact \texttt{svegetti@mpa-garching.mpg.de}. The methodology used for the lens
modelling is fully explained by \citet{powell_2021}.

\bmhead{Competing interests statement}  

The authors declare no competing interests.

\newpage

\section*{Tables}

\begin{table}[h]
\caption{{\bf Summary statistics for models of detection $\cal{V}$ that are the same redshift of the main lens.} Inferred parameters of the mass profile models tested for detection ${\cal V}$ that are at the same redshift of the main lens, $z=0.881$. $\Delta\ln{{\cal E}}$ is the Bayes factor relative to the model that fits the best over all those considered. This model is preferred over a model with no perturber by $\Delta\ln{{\cal E}_s}=388$. Models are listed in order of decreasing $\Delta\ln{{\cal E}}$. Here $M$ is the mass of the extended component within $r_{\rm max}$ for all NFW models and for the KG$_{\infty}$ model, and the total mass $M_{\rm t}$ for all other models. The characteristic radius $R_{\rm ch}$ is the effective radius of the extended component for all Sérsic models and for the uniform disk with a point-mass (UD+PM) and the Plummer with a point-mass (PLU+PM) models, the truncation radius for all the pseudo-Jaffe (PJ) models, the core radius for all the King models,  $r_{\rm max}$ for all NFW models, and the break radius for the broken power-law (bPL) models. M$_{\rm cyl,90}$ is the cylindrical mass within a radius of 90 pc. M$_{\rm PM}$ is the total mass of the point-mass component in
composite models.}
\begin{tabular}{@{}ccccccc@{}}
\toprule
Model & $\Delta\ln{\cal{E}}$ & M & M$_{\rm PM}$ & M$_{\rm cyl, 90}$ & R$_{\rm ch}$ & \\
&&[$10^{6} M_{\odot}$]& [$10^{5} M_{\odot}$] &[$10^{6} M_{\odot}$] & [pc] &\\
\midrule
UD+PM & 0 & 1.76 $\pm$ 0.10 & 4.25 $\pm$ 0.21 & 1.167 $\pm$ 0.039 & 98 $\pm$ 3 & $n \rightarrow 0$\\
\\
SER+PM & $-$3.5 & 1.79 $\pm $ 0.11 & 4.24 $\pm$ 0.20 & 1.175 $\pm$ 0.038 & 98 $\pm$ 3 & n = 0.071 $\pm$ 0.046\\ 
\\
bPL+PM  & $-$11 & 1.84 $\pm$ 0.12 & 3.70 $\pm$ 0.33 &  1.155 $\pm$ 0.043 & 142 $\pm$ 5 & $t_{\rm 1}=$ 0.41 $\pm$ 0.08\\
&&&&&& $t_{\rm 2}=$ 14 $\pm$ 3\\
\\
PLU+PM & $-$17 & 2.63 $\pm$ 0.25 & 3.80 $\pm$ 0.26 & 1.178 $\pm$ 0.043 &  136 $\pm$ 10 & - \\
\\
bPL & $-27$ & 2.12 $\pm$ 0.14 &-& 1.164 $\pm$ 0.039 & 193 $\pm$ 20 & $t_{\rm 1}=$ 1.29 $\pm$ 0.04\\
&&&&&& $t_{\rm 2}=$ 14 $\pm$ 4\\
\\
PJ+PM & $-$30 & 3.42 $\pm$ 0.43 & 1.93 $\pm$ 0.36 & 1.164 $\pm$ 0.033 &  266 $\pm$ 48 & -\\
\\
NFW+PM & $-$31 & 0.97 $\pm$ 0.17 & 3.03 $\pm$ 0.32 & 1.168 $\pm$ 0.037 & 145 $\pm$ 30 & $\log{c_{\rm V}}$ =  5.36 $\pm$ 0.20\\
\\
PL & $-$38 &-&-& 1.182 $\pm$ 0.033 &-& $\gamma$ = 2.330 $\pm$0.032 \\
\\
SER & $-$39 & 4.54 $\pm$ 0.67 & -& 1.197 $\pm$ 0.035 &  323 $\pm$ 84 & n = 7.39 $\pm$ 0.72\\
\\
PJ & $-$39 & 2.85 $\pm$ 0.27 &-& 1.228 $\pm$ 0.040 & 151 $\pm$ 19 & -\\
\\
PL+PM & $-$39 &-& 1.83 $\pm$ 0.42 & 1.169 $\pm$ 0.032 &-& $\gamma$ = 2.155 $\pm$ 0.060\\
\\
ePL+PM & $-$40 &-& 1.81 $\pm$ 0.42 & 1.190 $\pm$ 0.034 &-& $\gamma$ = 2.155 $\pm$ 0.056\\
&&&&&& $\theta$  = 95 $\pm$ 11 [deg]\\
&&&&&& $q$ = 0.58 $\pm$ 0.13\\
\\
NFW & $-$44 & 0.36 $\pm$ 0.03 &-& 1.197 $\pm$ 0.045 & 24 $\pm$ 4 & $\log{c_{\rm V}}$ = 7.25 $\pm$ 0.17\\
\\
KG$_{\infty}$ & $-$52 & 0.43 $\pm$ 0.04 &-& 1.165 $\pm$ 0.049 & 9.24 $\pm$ 1.18 & - \\
\\
KG & $-$53 & 3.05 $\pm$ 0.57 &-& 1.165 $\pm$ 0.050 &  9.39 $\pm$ 1.21  & $r_{\rm t}$ = (2.47 $\pm$ 2.24)$\times 10^4$ [pc]\\
\\
KG$_{\rm td}$ & $-$122 & 0.87 $\pm$ 0.04 &-& 0.871 $\pm$ 0.037 & 10 $\pm$ 5 & $r_{\rm t}$ =  40.89 $\pm$ 0.58 [pc]\\
\\
PM & $-$128 & 1.45 $\pm$ 0.07 &-&-&-&-\\
\\
NFW$_{\rm CDM}$ & $-$147 & 0.81 $\pm$ 0.03 &-& 1.182 $\pm$ 0.042 & 85 $\pm$ 6 & $\log{c_{\rm V}}$ = 5.96 $\pm$ 0.07\\
\botrule
\end{tabular}
\label{tab:main_tab1}
\end{table}

\begin{table}[h]
\caption{{\bf Summary statistics for models of detection $\cal{V}$ where the redshift is a free parameter.}  $\Delta\ln{{\cal E}}$ is the Bayes factor relative to the model that fits best over all those considered, that is, the UD+PM in Table \ref{tab:main_tab1}.}
\begin{tabular}{@{}cccccccc@{}}
\toprule
Model & $\Delta\ln{\cal{E}}$ & M$_{\rm vir}$ & M$_{\rm PM}$ & M$_{\rm cyl, 90}$ & & z \\
&&[$10^{6} M_{\odot}$] &[$10^{5} M_{\odot}$] & [10$^{6} M_\odot$] & &\\
\midrule
NFW$_{\rm field}$+PM & $-$34 & 5.63 $\pm$ 0.75 & 3.12 $\pm$ 0.35 & 1.209 $\pm$ 0.051 &$\log{c_{\rm vir}}=$ 1.68 $\pm$ 0.08 & 0.925 $\pm$ 0.059\\
\\
PL$_{\rm field}$ & $-$40 &-&-& 1.632 $\pm$ 0.211 & $\gamma$ = 2.405 $\pm$ 0.043 & 1.32 $\pm$ 0.16\\
\\
NFW$_{\rm field}$ & $-$40 & 2.86 $\pm$ 0.21 &-& 1.516 $\pm$ 0.086 & $\log{c_{\rm vir}}=$ 2.91 $\pm$ 0.09 & 1.469 $\pm$ 0.046\\
\\
PL$_{\rm field}$+PM & $-$41 &-& 1.82 $\pm$ 0.45 &  1.241 $\pm$ 0.089 & $\gamma$ = 2.202 $\pm$ 0.055 & 0.983 $\pm$ 0.084\\
\\
NFW$_{\rm CDM, field}$ & $-$75 & 4.04 $\pm$ 0.27 &-& 1.226 $\pm$ 0.039 & $\log{c_{\rm vir}}=$ 2.18 $\pm$ 0.04 & 0.883 $\pm$ 0.028\\
\botrule
\end{tabular}
\label{tab:main_tab2}
\end{table}

\clearpage

\section*{Figures}

\setcounter{figure}{0}
\renewcommand{\figurename}{Figure}

\begin{figure}[h]
\centering
\includegraphics[width=0.7\textwidth]{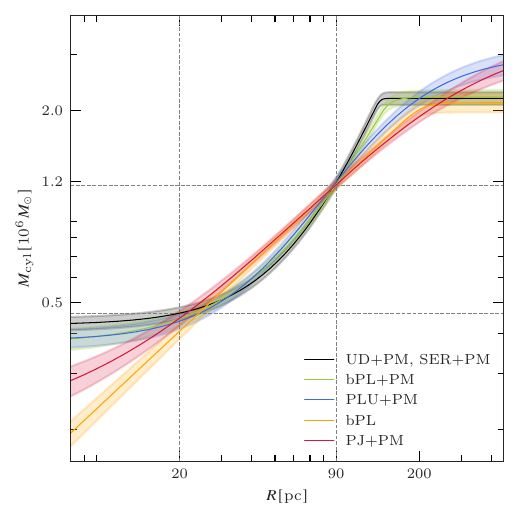}
\caption{{\bf The cylindrical mass profile ($M_{\rm cyl}$) for the six best-fitting models at $z=z_{\rm lens}$.} The vertical lines represent the 20 and 90 pc radii, which is also where the different models most closely agree with each other. The horizontal lines are the corresponding values of $M_{\rm cyl}$ for the uniform disk and point mass (UD+PM) model. In the legend, models appear in order of decreasing Bayes factor. Note that the top two models, a uniform disk with a point mass and a S\'ersic profile with a point mass (denoted UD+PM and SER+PM, respectively), are plotted as a single curve and uncertainty band since the differences between them are too small to be easily distinguished at this plotting scale. Also shown are the cylindrical mass profiles for a broken power-law with a point mass (bPL+PM), a Plummer sphere with a point mass (PLU+PM), a broken power-law (bPL) and a pseudo-Jaffe with a point mass (PJ+PM). The uncertainty bands represent the 1-$\sigma$ confidence interval around the mean.}  
\label{fig:main_fig1}
\end{figure}

\clearpage

\section*{Extended Data}

\captionsetup[table]{name=Extended Data Table}
\renewcommand{\figurename}{Extended Data Figure}
\setcounter{figure}{0}
\setcounter{table}{0}

\begin{figure}[h]
\centering
\includegraphics[width=0.5\textwidth]{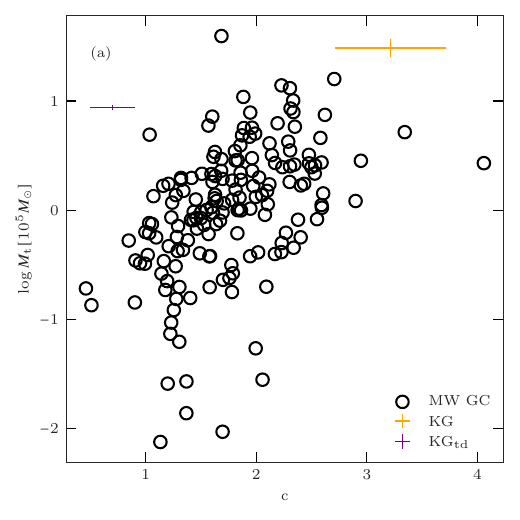}
\hspace{-0.2cm}
\includegraphics[width=0.5\textwidth]{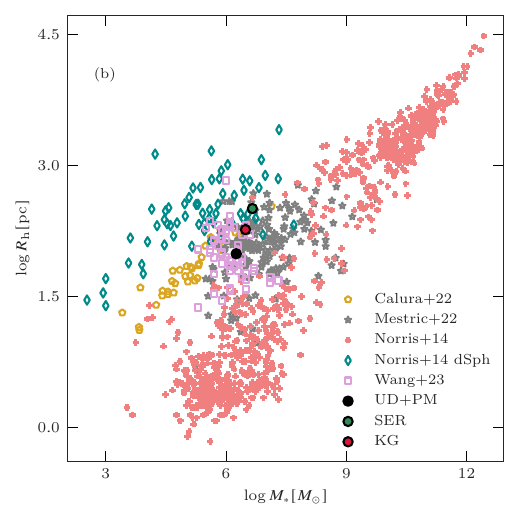}
\caption{{\bf Comparison with known stellar systems.} Panel (a): Mass and concentration for the GCs in the Milky Way \citep[black circles,][]{Baumgardt_2023} and for the KG$_{\rm{td}}$ and KG models of detection ${\cal V}$ (coloured points, mean value and 1-$\sigma$ uncertainty). Panel (b): relation between stellar mass and effective radius from observations of different types of objects: dSph galaxies at $z = 0$ \citep[green diamonds,][]{norris_2014},  strongly lensed star-forming clumps at $z = 2$ to 6 \citep[gray stars,][]{mestric_2022}, eUCDs in the Virgo cluster \citep[pink squares,][]{wang_2023}, dEs, dS0s, nuclear star clusters, GCs, UCDs, cEs and YMCs at $z = 0$ \citep[coral crosses,][]{norris_2014}. The yellow diamonds represent simulated star-forming stellar clusters at $z\sim 6$. The green, red and black dots are detection ${\cal V}$ when modelled with a Sérsic profile of index $n\sim7$, a King profile and a Sérsic profile of index $n\rightarrow 0$ with a point-mass, respectively.}
\label{fig:ed_fig1}
\end{figure}

\begin{figure}[h]
\centering
\includegraphics[width=0.8\textwidth]{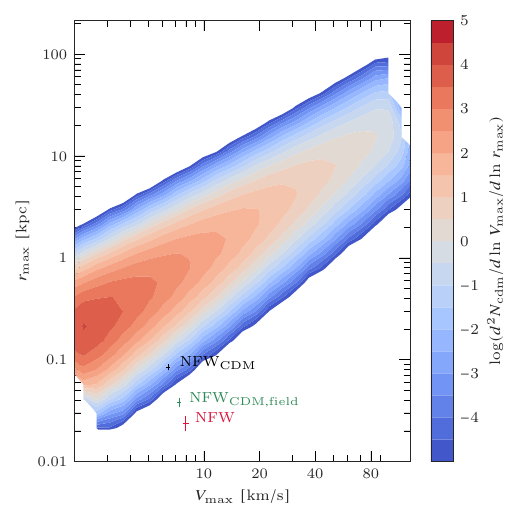}
\caption{{\bf Comparison with CDM predictions.} The contours represent the number density of CDM subhaloes with $M_{\rm{sub}}>10^5 h^{-1} M_\odot$. These were obtained using the SASHIMI-C semi-analytical subhalo model \citep{hiroshima_2018, ando_2019} assuming a host redshift and mass of $z_{\rm h} = 0.881$ and  $M_{\rm h} = 10^{12}M_\odot$. The coloured crosses show the $V_{\rm max}$ and $r_{\rm max}$ mean values and 1-$\sigma$ uncertainties inferred for different models of the detected object: an NFW profile with free concentration (NFW,  $\Delta\ln{\cal E} =-44$), an NFW halo with concentration drawn from the mass-concentration relation in \citep{duffy_2008} (NFW$_{\rm CDM, field}$, $\Delta\ln{\cal E} =-75$) and an NFW subhalo with concentration drawn from \citep{moline_2023} (NFW$_{\rm CDM}$, $\Delta\ln{\cal E} =-147$).} 
\label{fig:ed_fig2}
\end{figure}

\begin{figure}[h]
\centering
\includegraphics[width=0.7\textwidth]{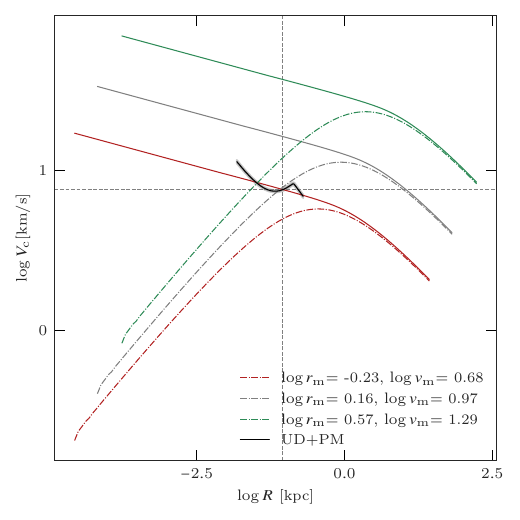}
\caption{{\bf Comparison with SIDM predictions.} Projected ``circular velocity'' versus projected radius for different SIDM haloes (solid lines), for their CDM NFW analogues (dashed-dotted lines) and for our best-fit model, a uniform disk with a point mass (UD+PM). The vertical and horizontal lines represent the 90 pc radius at which all models for the detected object agree most closely, and the corresponding circular velocity. The NFW profiles are labelled according to their log values of $V_{\rm max}$ and $r_{\rm max}$. The uncertainty band in the UD+PM curve represent the 1-$\sigma$ error around the mean value.} 
\label{fig:ed_fig3}
\end{figure}

\begin{table}[h]
\caption{{\bf Macro-model parameters} Inferred macro-model parameters, including the mass and position of detection ${\cal A}$, for the best-fit case where detection $\cal{V}$ is modelled as a uniform disk with a point-mass (UD+PM).}
\label{tab:macro}%
\begin{tabular}{@{}cccccccccccc@{}}
\toprule
$\kappa_0$            & 0.622 $\pm$ 0.003\\
$\theta$ [deg]        & $-$18.6 $\pm$ 0.2\\
$f$                   & 0.811  $\pm$ 0.002\\
$x$ 	[arcsec]      & $-$0.1234 $\pm$ 0.0003\\
$y$ 	[arcsec]      & $-$0.1187 $\pm$  0.0004\\
$\gamma$              & 1.747  $\pm$ 0.006\\
$\Gamma$              & 0.0494 $\pm$ 0.0008\\
$\Gamma_\theta$ [deg] & 80.5  $\pm$ 0.2\\
$a_3$                 & $-$0.0046  $\pm$ 0.0003\\
$b_3$                 & 0.0025   $\pm$ 0.0002\\
$a_4$                 & $-$0.0004 $\pm$ 0.0001\\
$b_4$                 & $-$0.0044 $\pm$ 0.0002\\
$M_{\rm t, {\cal A}}$ [10$^7 M_\odot$] &  7.83 $\pm$  1.13\\
$x_{{\cal A}}$ [arcsec] & $-$0.123 $\pm$ 0.001\\
$y_{{\cal A}}$ [arcsec] & 0.456$\pm$  0.004\\
\botrule
\end{tabular}
\label{tab:ed_tab1}
\end{table}

\renewcommand{\figurename}{Supplementary Figure}
\setcounter{figure}{0}

\clearpage

\section*{Supplementary Information}
\label{sec:supplementary}

\subsection*{Comparison with dark matter models}
\label{dm_comparison}

We have tested 23 models for the mass density profiles of detection ${\cal V}$. Here, we discuss in more detail the implications of its inferred structure for the nature of dark matter, assuming that it is a dark matter-dominated object. Given the area of sky probed by the data, \citet{powell_2025} have estimated the probability of detecting at least one subhalo with a mass between 10$^6$ M$_\odot$ and 10$^7$ M$_\odot$ to be 0.65 and 0.14,  for CDM and for WDM with $m_{\rm wdm} = 4.6$ keV, respectively. Hence, detecting such an object is not {\it a priori} unlikely. However, we find that in CDM or WDM models, its predicted structure is incompatible with that inferred for our VLBI detection (see Extended Data Figure  \ref{fig:ed_fig2}).

The NFW$_{\rm field}$ halo has a fitted concentration of $\log c_{\rm vir}= 2.91\pm0.09$. A CDM NFW halo with a mass and redshift consistent with our analysis ($M_{\rm vir}\sim 2.9\times 10^6 M_\odot$ and $z\sim1.4$) is expected to have a virial concentration of $\log c_{\rm vir}=1.10\pm0.15$ \citep{duffy_2008} or $\log c_{\rm vir} = 1.04 \pm 0.09$ \citep{ludlow_2016}, depending on the assumed concentration-mass relation. Hence, the mean log inferred value is 10 or 14$\sigma$ larger than the median of the lognormal distributions predicted by N-body simulations; here $\sigma$ is the sum in quadrature of the uncertainty on the inferred value and the scatter in the lognormal relations. Similarly, the NFW subhalo model has a $\log c_{\rm v} = 7.25\pm0.17$, which is 20$\sigma$ larger than the median value of $\log c_{\rm v}=3.90\pm0.02$ \citep{moline_2023} predicted for a subhalo with a peak circular velocity of $V_{\rm max} = 7.98$ km s$^{-1}$ and peak radius r$_{\rm max} = 24$ pc at  z = 0.881.

To quantify the statistical level at which the CDM model is incompatible with the data, we consider two extra models, NFW$_{\rm CDM, field}$ and NFW$_{\rm CDM}$. The former has a free redshift and $c_{\rm vir}$ drawn from a lognormal distribution with median and scatter taken from numerical simulations \cite[specifically from ref.][]{duffy_2008}. The second model assumes the perturber to be at the same redshift as the lens, and its $r_{\rm max}$ is drawn from a lognormal distribution with median and scatter taken from \cite{moline_2023}. The NFW$_{\rm CDM, field}$ model prefers a halo at z = 0.883 $\pm$ 0.028 (this is consistent with the lens redshift within uncertainties) with $M_{\rm vir}=\left(4.04 \pm 0.27\right)\times 10^6 M_\odot$ and $\log c_{\rm vir}$ = 2.18 $\pm$ 0.04, but with $\Delta\ln{\cal{E}}=-75$. The subhalo model NFW$_{\rm CDM}$ gives a best fit for a subhalo with M$_{\rm max}=\left(8.13 \pm 0.27\right)\times 10^5 M_\odot$ within $r_{\rm max} = 85.0 \pm 5.7$ pc, but with $\Delta\ln{\cal{E}}=-147$. In both cases, the inferred concentrations are very far out in the high-concentration tail of the predicted distributions, leading to the models being very strongly disfavoured, at a level between 12 and 17$\sigma$. We note that the relatively low redshift inferred for NFW$_{\rm CDM, field}$ is due to a combination of the redshift dependence of the mass concentration relation and the fact that most of the prior volume is at relatively low concentration values; that is, perturbers at lower redshift can provide a similar lensing effect to higher redshift ones for a smaller and hence more probable concentration (Tajalli et al. in prep.).

Once a point mass is added to the NFW model, we infer smaller concentrations for the extended component, $\log{c_{\rm V}}$ =  5.36 $\pm$ 0.20 at $z\equiv z_{\rm lens}$ and $\log{c_{\rm vir}} = 1.69 \pm 0.08$ at $z = 0.925 \pm 0.059$. These values are nominally 6 and 3$\sigma$ larger than the predicted values for dark-matter-only objects of the same mass and redshift, but a direct comparison to such predictions is not appropriate. This is because a central black hole would substantially increase the concentration of a dark matter halo, reducing or eliminating the conflict with our observational data, and more importantly, because our inferred black hole masses are much larger than is plausible in CDM haloes of such low mass, which, in addition, are not even expected to form stars. Ultimately, however, all variations of the NFW models are highly disfavoured by the data when compared to the best-fitting UD+PM model. Hence, we conclude that if our detection is confirmed to be dark-matter dominated, it will rule out CDM at a statistical level between 8 and 17$\sigma$, and WDM with even higher significance, increasing with decreasing particle mass.

Existing gravothermal fluid models \citep[e.g.][]{outmezguine_2023}, semi-analytical models \citep[e.g.][]{yang_2024,ando_2024} and N-body simulations \citep[e.g.][]{turner_2021} indicate that SIDM haloes can have density profiles with super-isothermal slopes in their inner regions once they are deep in the core collapse phase of their evolution. To investigate whether the properties of detection ${\cal V}$ are consistent with SIDM expectations, we focus on the gravothermal fluid model by ref. \cite{outmezguine_2023}. Specifically, we fit the mass density profile at the latest available time for their $n=3.7$ model (their Figure E1) and derive from it the corresponding projected ``circular velocity'' profile, defined as $V_{\rm c}(x)/V_{\rm max}=0.61\sqrt{M_{\rm cyl}(x)/(\rho_{\rm s}r_{\rm s}^3 x)}$, with $x = R/r_{\rm s}$. Here $\rho_{s}$, $r_{\rm s}$ and $V_{\rm max}$ should be understood to be the parameters of the NFW halo in the absence of dark-matter interactions. Since fluid models, such as the one considered here, follow the evolution of idealised haloes decoupled from their cosmological context, these NFW equivalent parameters are independent of time. This is in contrast to ``real'' CDM haloes that grow substantially in time through accretion and merging. However, we know of no cosmological SIDM  simulations with spatial resolution adequate for comparison with the VLBI data. Hence, we cannot determine how the structure of an SIDM halo with mass as small as detection ${\cal V}$ has been affected by the fact that it has assembled progressively with time rather than being present, effectively in isolation, since some early epoch. We therefore take $V_{\rm max}$ and $r_{\rm max}$ (equivalently $\rho _{\rm s}$ and $r_{\rm s}$) to be the values for a z = 0.881 CDM  analogue, which we assume to have evolved in isolation for the Hubble time at redshift $z$. This is likely to overestimate the structure modification due to self-interactions, so the cross-sections needed to match the observation can be taken as lower limits on those needed in a more realistic halo assembly model. 

Our best-fitting model, UD+PM, is most robustly constrained at a radius of 90 pc, with $V_{\rm c}(90~{\rm pc}) = 7.47 \pm 0.12 $ km s$^{-1}$. Extended Data Figure  \ref{fig:ed_fig3} shows circular velocity profiles for SIDM haloes with differing masses, together with those of their CDM analogues. The mass-concentration relation of the latter is assumed to be given by the median estimated by ref. \cite{duffy_2008} at z = 0.881. We find that an SIDM profile will have $V_{\rm c}$ at 90 pc matching the observed value if its CDM analogue has $V_{\rm max}=4.80$ km s$^{-1}$ and $r_{\rm max} = 0.60$ kpc.  Assuming this halo has been evolving for the age of the Universe at $z=0.881$ and adopting the normalising time-scale by ref. \cite{outmezguine_2023} gives a lower limit on the effective cross-section of $\sigma_{\rm c,0}/m_{\rm dm} = 796$ cm$^2$ g$^{-1}$. We caution, however, that this number is proportional to $V_{\rm max}^{-3}r_{\rm max}^2$ and so is very sensitive to the assumed $V_{\rm max}$--$r_{\rm max}$ relation and its scatter. 

Looking more closely at Extended Data Figure  \ref{fig:ed_fig3}, it can be seen that the theoretical SIDM profile is not a good match to that inferred from the observations; its mean slope over the observationally well-determined range is significantly shallower. Soon after the time shown, the SIDM model will form a black hole, which will then grow rapidly through Bondi accretion, potentially reaching masses similar to that of the central component of our UD+PM model. Unfortunately, however, there are no reliable calculations of this initial black hole growth phase, and so, no detailed predictions of the post-collapse $M_{\rm cyl}(R)$ that we can compare with the data. The discrepancy in slope between our observations and the predictions from ref. \cite{outmezguine_2023} may be related to the fact that their profile is that of an idealised isolated SIDM halo that is still in the long-mean-free-path stage of its evolution. \citet{Gad-Nasr_2024} have simulated even later (but still pre-collapse) phases, where the inner regions are deep into the short-mean-free-path regime. Their work offers some indication that it may be possible to achieve profile shapes similar to those required observationally for appropriately chosen velocity-dependent cross-sections. Verifying this must await simulations, which can follow evolution through black hole formation and the subsequent Bondi accretion phase. Only then will it be possible to tell whether objects with the structure observed for detection $\cal{V}$ can form naturally in a SIDM-dominated Universe, and, if so, what parameters this requires for the scattering cross-section.

\subsection*{Comparison with GCs and UCDs}
\label{gc_comparison}

In the mass range relevant for our main lens, the number of globular clusters (GCs) in a galaxy is expected to be significant and, in some cases, comparable to the number of dark matter subhaloes and field haloes \citep{he_2018}. The lens galaxy in the gravitational lens system JVAS B1938+666 is an early-type that has a stellar mass within the Einstein radius of 4.7$\times10^{10} M_\odot$ \citep{lagattuta_2012}, which implies a total number of GCs around 250 \citep{he_2018}. Assuming that the spatial density of the GCs follows a Sérsic profile with the same index and effective radius as the host galaxy \citep{saifollahi_2025} and adopting the mass function by \citet{jordan_2009}, we estimate a density of about 100 GCs per square arcsec in an annulus of 5 milliarcseconds around the lensed images. As a comparison, the expected number density of subhaloes within the same area and between (10$^6$ to 10$^9 M_\odot$) is $\left(9.46 \pm 0.04\right)\times10^5$ f$_{\rm sub}^{-1}$ arcsec$^{-2}$ or $\left(1.135 \pm 0.005\right)\times 10^4$ per square arcsec for a dark matter fraction in substructure of 0.012 \citep{hsueh_2020}, where the error reflects the uncertainty on the total lens mass. Given the lens and source redshifts, field haloes are expected to be roughly an order of magnitude more numerous \citep{despali_2018}. Only a fraction of haloes and subhaloes are expected to core-collapse and so be detectable through their lensing effect. Hence, the relative probability of detecting a dark matter halo or a GC and an ultracompact dwarf (UCD) will depend on the details of the dark-matter model, on the as-yet unknown fraction of dark matter in subhaloes, as well as on the exact sensitivity of the data to the presence of each population. 

The KG models we use to describe typical GCs are all strongly disfavoured. Those highest in the ranking are KG$_{\infty}$ and KG, which have a very similar $\Delta\ln{\cal{E}}$ of $-52$ and $-53$, respectively. The characteristics of these two models that are best constrained are M$_{\rm cyl, 90}\simeq10^6$ M$_{\odot}$ and r$_{\rm c}\simeq10$ pc. Although rare, Milky Way GCs exhibiting similar properties exist, the closest example being NGC~2419, a GC located in the distant Galactic halo \citep[][]{ripepi07}. In fact, according to the Baumgardt catalogue of Milky Way GCs, NGC~2419 has a current mass of $\sim8\times10^5$ M$_{\odot}$ (and an initial mass, decreased by internal dynamical processes, of $1.4\times10^6$ M$_{\odot}$) and a core radius of 8.3 pc. The KG$_{\rm td}$ model is even more strongly disfavoured, and has a concentration that is very small for a GC of its mass (see Extended Data Figure  \ref{fig:ed_fig1}, panel (a)). Massive clusters tend to have shorter internal dynamical timescales (\citep[see e.g.,][]{harris_2010}), which favours evolution of the cluster structure towards higher concentration (as shown by the trend in panel (a) of Extended Data Figure  \ref{fig:ed_fig1}). To be this massive yet with such low concentration, a cluster would have to be born already extended \citep[lower stellar density implies longer dynamical timescales,][]{meylan97}. The closest example to this case is once again NGC~2419, which has the longest relaxation time at the half-light radius \citep[][]{harris_2010} among all Milky Way GCs. Our KG$_{\rm td}$ model is even less concentrated than this exceptional cluster.

It is worth noting that a recent exploration of resolved Milky Way GCs has confirmed that the structural parameters recovered with the unprecedentedly deep {\it Euclid} photometry are in good agreement with previous studies based on brighter, more massive stars. Even including stars with a mass as low as 0.16 M$_\odot$, deviations from standard King models only become mildly relevant in the outermost regions, close to the cluster tidal radius \citep{massari_2025}. This can affect the size of the tidal radius only by up to 20\%. Given all these considerations and the fact that the KG$_{\rm td}$ and KG models are both strongly disfavoured by the data (although at very different levels), we deem it highly unlikely that the object under study is a GC, even if its recovered properties are not completely out of the range of those observed in the Milky Way.

UCDs were discovered in the early 2000s \citep{hilker_1999, phillipps_2001, drinkwater_2003} as a new class of object whose properties lie between those of GCs and dwarf galaxies: their effective radii span an approximate range between 10 and 100 pc, they have stellar masses between 10$^6$ and 10$^8~M_\odot$, and typical luminosities of $L\sim10^7 L_\odot$ \citep[e.g.][]{hasegan_2005, mieske_2008, misgeld_2011, norris_2014}. They are some of the most compact stellar systems currently known, and even though they were initially discovered in galaxy clusters, they have since been observed in lower-density environments also \citep{norris_2011}. Different mechanisms have been proposed for the formation of different subgroups of UCDs \citep[e.g.][]{norris_2011, darocha_2011, chilingarian_2011}. According to one of these scenarios, some UCDs arise from the tidal stripping of dwarf galaxies; what is observed today is then the remnant stellar core of the parent object \citep[e.g.][]{bekki_2003,drinkwater_2003,gregg_2003,goerdt_2008,pfeffer_2014,paudel_2023}. This hypothesis is supported by observations in the local Universe of UCDs with extended envelopes (eUCDs) and tidal features, as well as with a central super-massive black hole. Single-component profiles, such as a King, a Sérsic or a de Vaucouleurs profile, describe well the light distribution of regular UCDs. However, in the case of eUCDs, they assume extreme parameters, such as poorly bounded tidal radii and very large concentrations ($\sim 3$) for a King profile, and large indices of around 8 for a Sérsic profile \cite{wang_2023}. This behaviour is similar to that displayed by our KG and SER models. As shown in panel (b) of Extended Data Figure  \ref{fig:ed_fig1}, if we assume that the total mass of the SER model is in stars, the object under study has a size and mass which lie on the edge of the distribution of these properties for eUCDs in the Virgo Cluster.

Our profile modelling for detection  ${\cal V}$  strongly supports an additional unresolved central component, however, and such a structure is much more compatible with the properties of a subset of UCDs that contain a compact nuclear star cluster (NSC). For example, our top model in the ranking (UD+PM) has a mass and a size entirely consistent with the nucleated UCD population in the Virgo cluster (see the black symbol in panel (b) of Extended Data Figure  \ref{fig:ed_fig1}). Moreover, the inferred ratio between the masses of the two components is consistent with  observations of the inner regions of at least some UCDs hosting a NSC \cite{graham_2025}. The starkest inconsistency between the properties of our favoured model and those observed in UCDs is the sharp mass cut-off beyond $90$ pc, as shown in panel (a) of Supplementary Figure \ref{fig:si_fig1}. This is challenging on astrophysical grounds, but could be due to the limitations of the macro-model for the main lens on these small angular scales, or to some unmodelled line-of-sight process affecting the radio data. The latter seems less likely given the observed surface brightness distribution of the gravitational arc. UCDs and eUCDs have so far been observed only within galaxy clusters, which may also disfavour this interpretation. 
Ultimately,  only a meaningful limit on the luminosity of detection $\cal{V}$ could provide a more definitive indication of its nature. If no stars were to be detected, it cannot be a UCD, since such objects are gravitationally dominated by their stars. However, this is beyond the reach of current optical and near-IR observing facilities.

\subsection*{Cylindrical mass profile}
\label{m_cyl}

Supplementary Figure \ref{fig:si_fig1} displays the cylindrical mass profiles for most tested models.

\subsection*{Posterior distributions}
\label{post_pdf}

Supplementary Figures \ref{fig:si_fig2} to \ref{fig:si_fig24} display the posterior distribution for the parameters of all the mass density profiles tested for detection $\cal{V}$. Note that in most of these plots, we use the projected mass within 90 pc as the amplitude parameter because it is very well constrained and almost independent of the other parameters.

\begin{figure}
\centering
\includegraphics[width=0.45\textwidth]{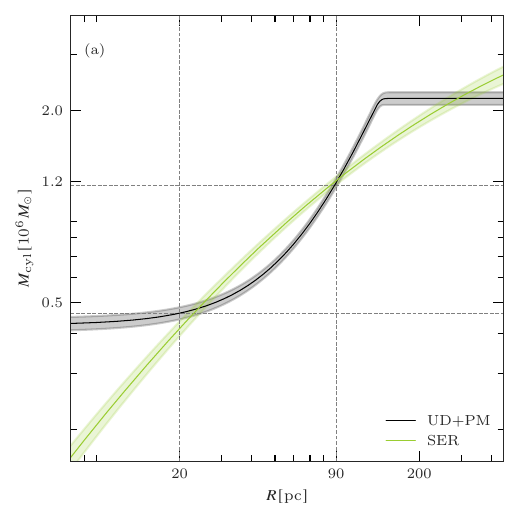}
\hspace{-0.3cm}
\includegraphics[width=0.45\textwidth]{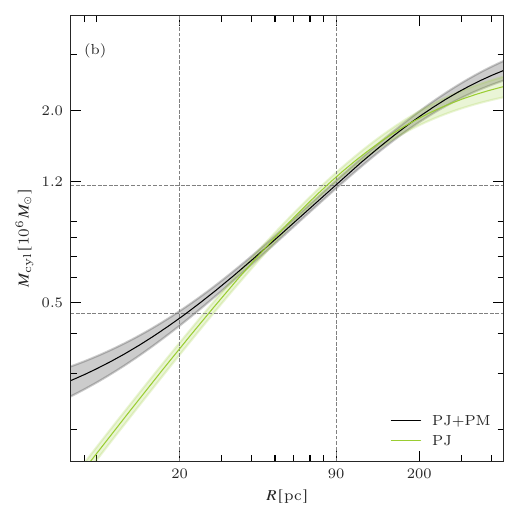}
\includegraphics[width=0.45\textwidth]{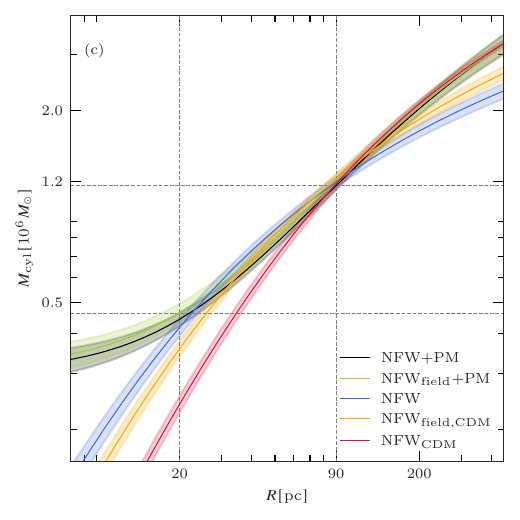}
\hspace{-0.3cm}
\includegraphics[width=0.45\textwidth]{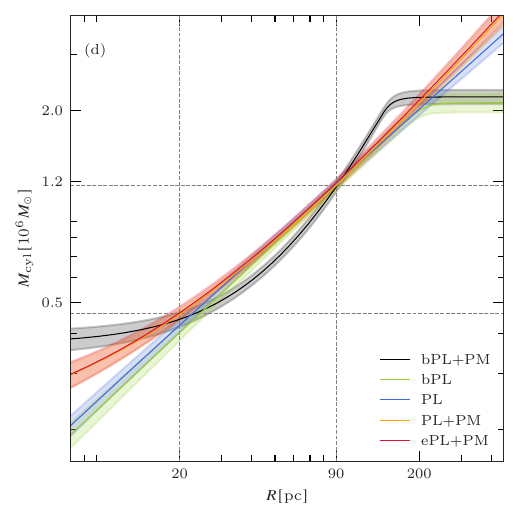}
\includegraphics[width=0.45\textwidth]{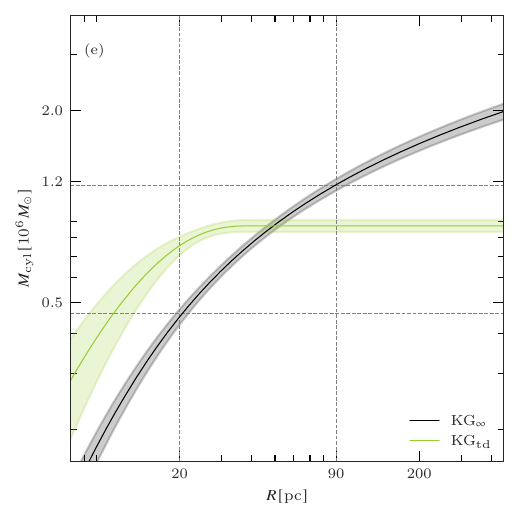}
\hspace{-0.3cm}
\includegraphics[width=0.45\textwidth]{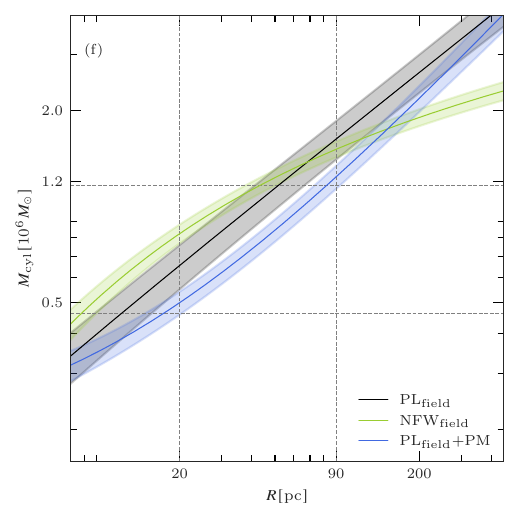}
\caption{{\bf Cylindrical mass profiles.} Panel (a): cylindrical mass profiles for the uniform disk plus point mass and Sérsic models. Panel (b): cylindrical mass profiles for the Pseudo-Jaffe models. Panel (c):  cylindrical mass profiles for the NFW models with a redshift consistent with that of the main lens. Panel (d): cylindrical mass profiles for the power-law models. Panel (e):  cylindrical mass profiles for the King models. We do not plot the KG model because it is indistinguishable from the KG$_{\rm \infty}$ one at this plotting scale. Panel (f): cylindrical mass profiles for the profiles with the redshift as a free parameter. In all panels, the vertical lines represent the 20 and 90 pc radii, which are also where the different models agree most closely. The horizontal lines are the corresponding values of $M_{\rm cyl}$ for the UD+PM model. In the legends, models appear in order of decreasing Bayes factor. The uncertainty bands represent the 1-$\sigma$ confidence interval around the mean.}  
\label{fig:si_fig1}
\end{figure}

\begin{figure}[h]
\includegraphics[width=1.4\textwidth, angle=90]{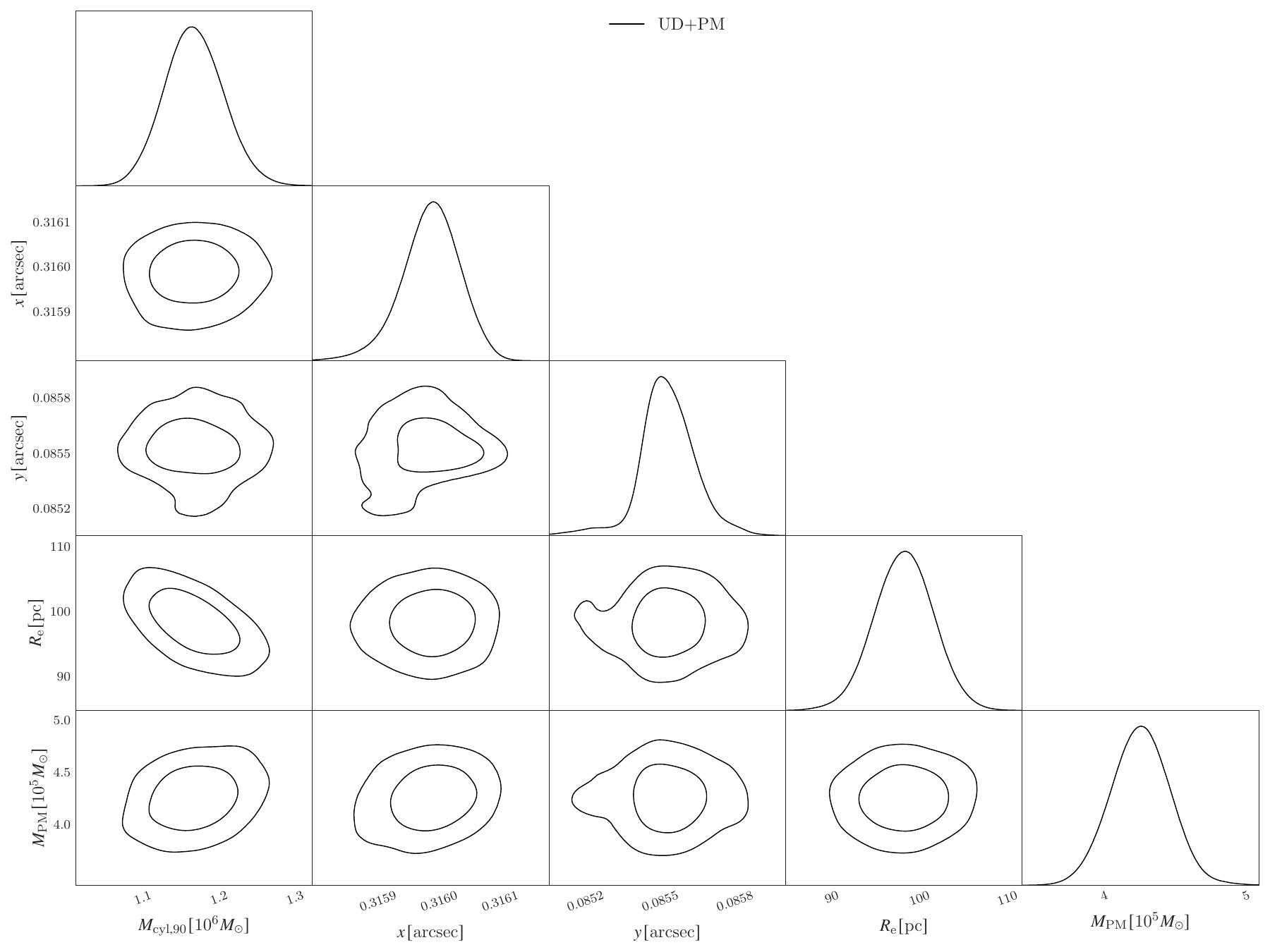}
\caption{{\bf Posterior distributions for the parameters of the uniform disk plus point-mass model.} The contours represent the 1- and 2-$\sigma$ confidence regions.} 
\label{fig:si_fig2}
\end{figure}

\begin{figure}[h]
\includegraphics[width=1.4\textwidth, angle=90]{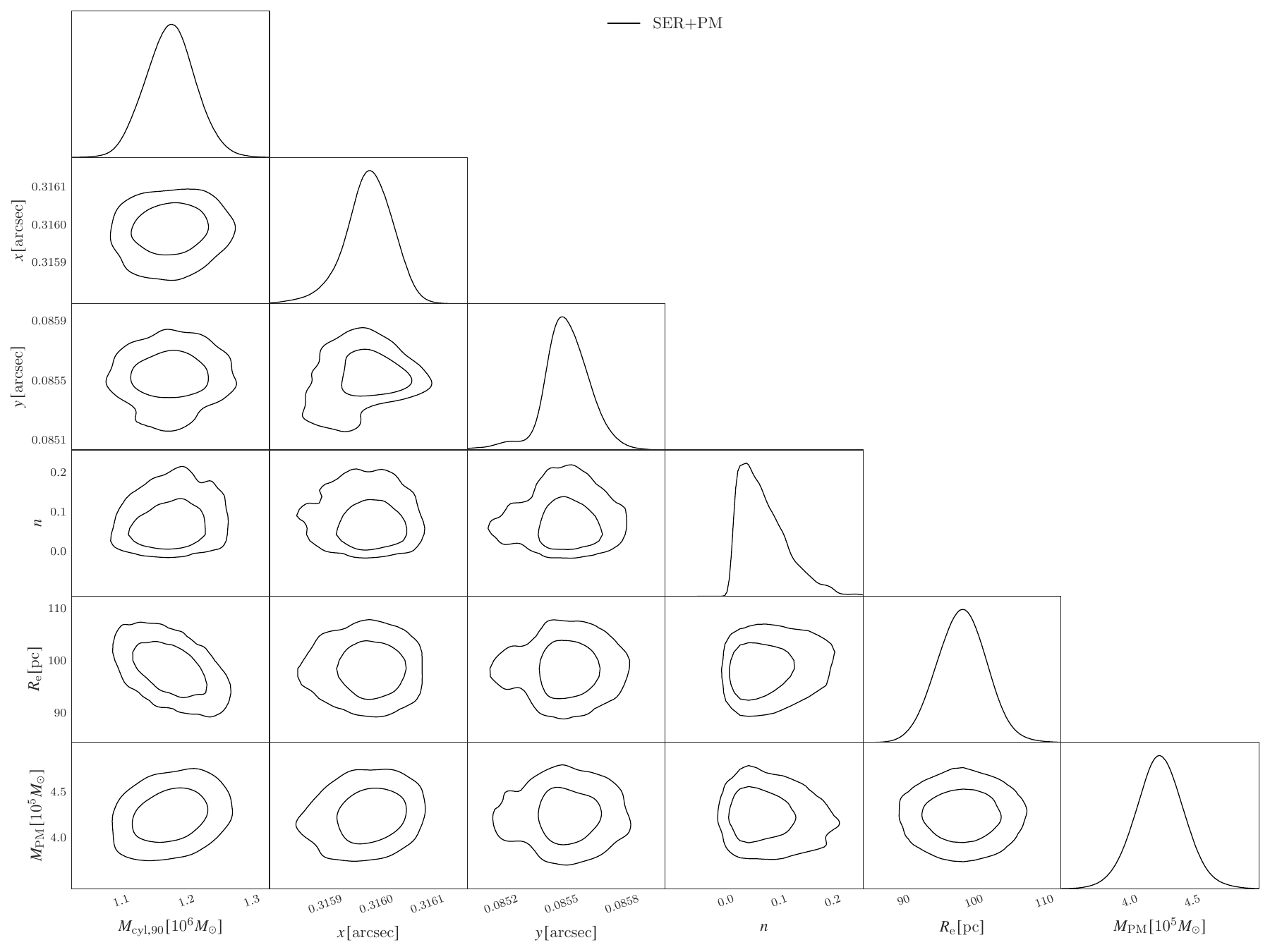}
\caption{{\bf Posterior distributions for the parameters of the Sérsic plus point-mass model.} The contours represent the 1- and 2-$\sigma$ confidence regions.} 
\end{figure}

\begin{figure}[h]
\includegraphics[width=1.4\textwidth, angle=90]{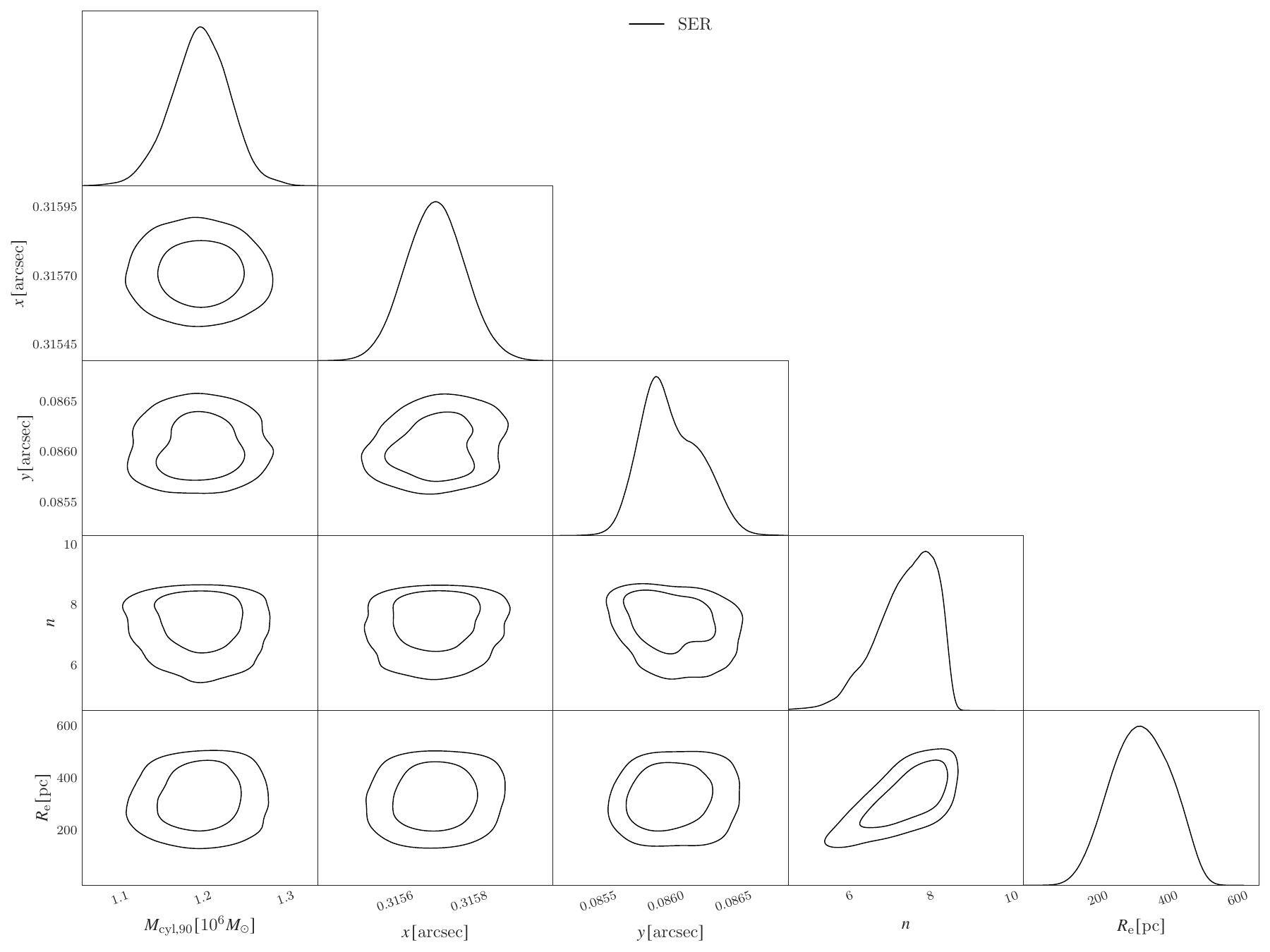}
\caption{{\bf Posterior distributions for the parameters of the Sérsic model.} The contours represent the 1- and 2-$\sigma$ confidence regions.} 
\end{figure}

\begin{figure}[h]
\includegraphics[width=1.4\textwidth, angle=90]{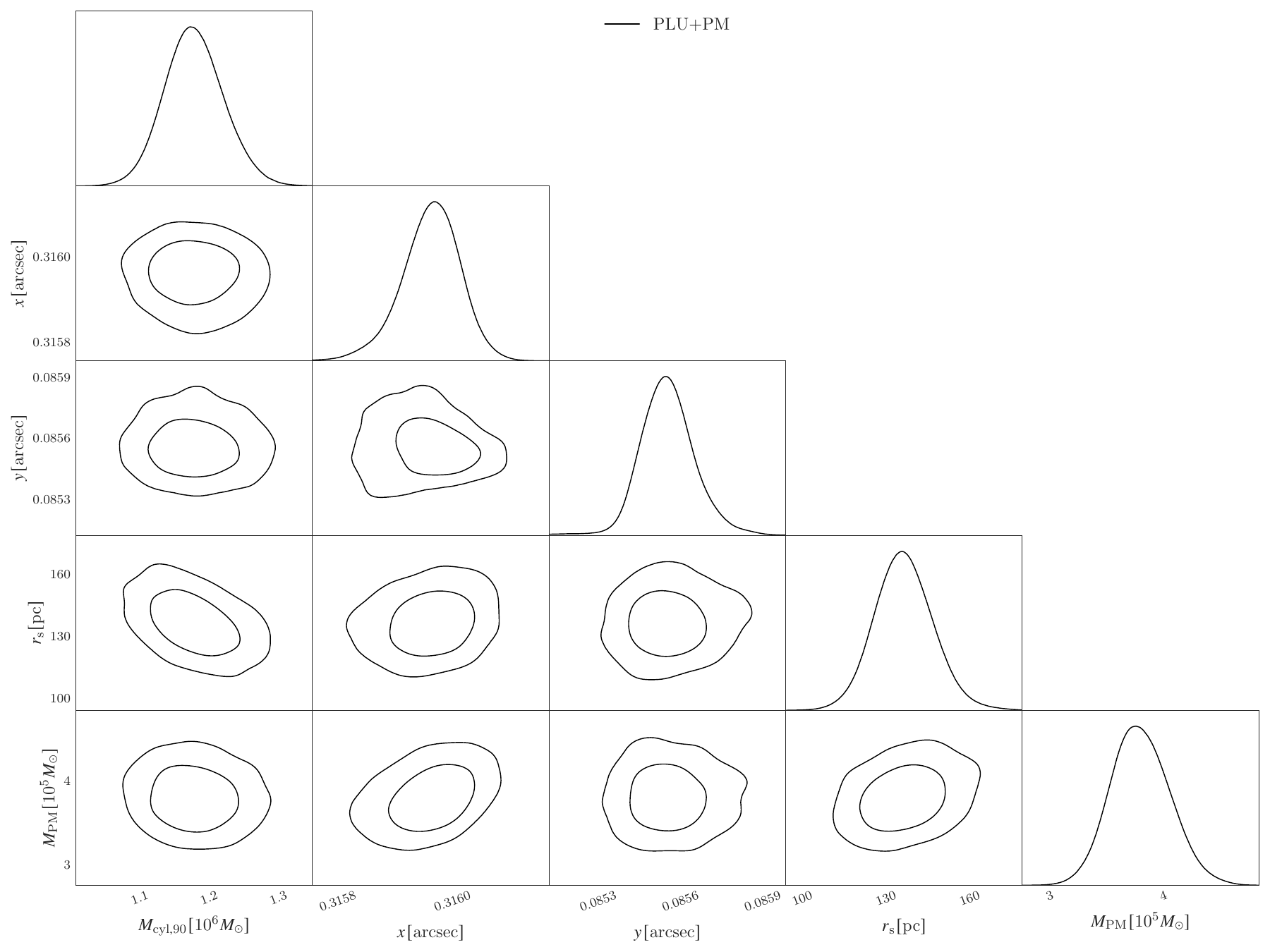}
\caption{{\bf Posterior distributions for the parameters of the Plummer plus point-mass model.} The contours represent the 1- and 2-$\sigma$ confidence regions.} 
\end{figure}

\begin{figure}[h]
\includegraphics[width=1.4\textwidth, angle=90]{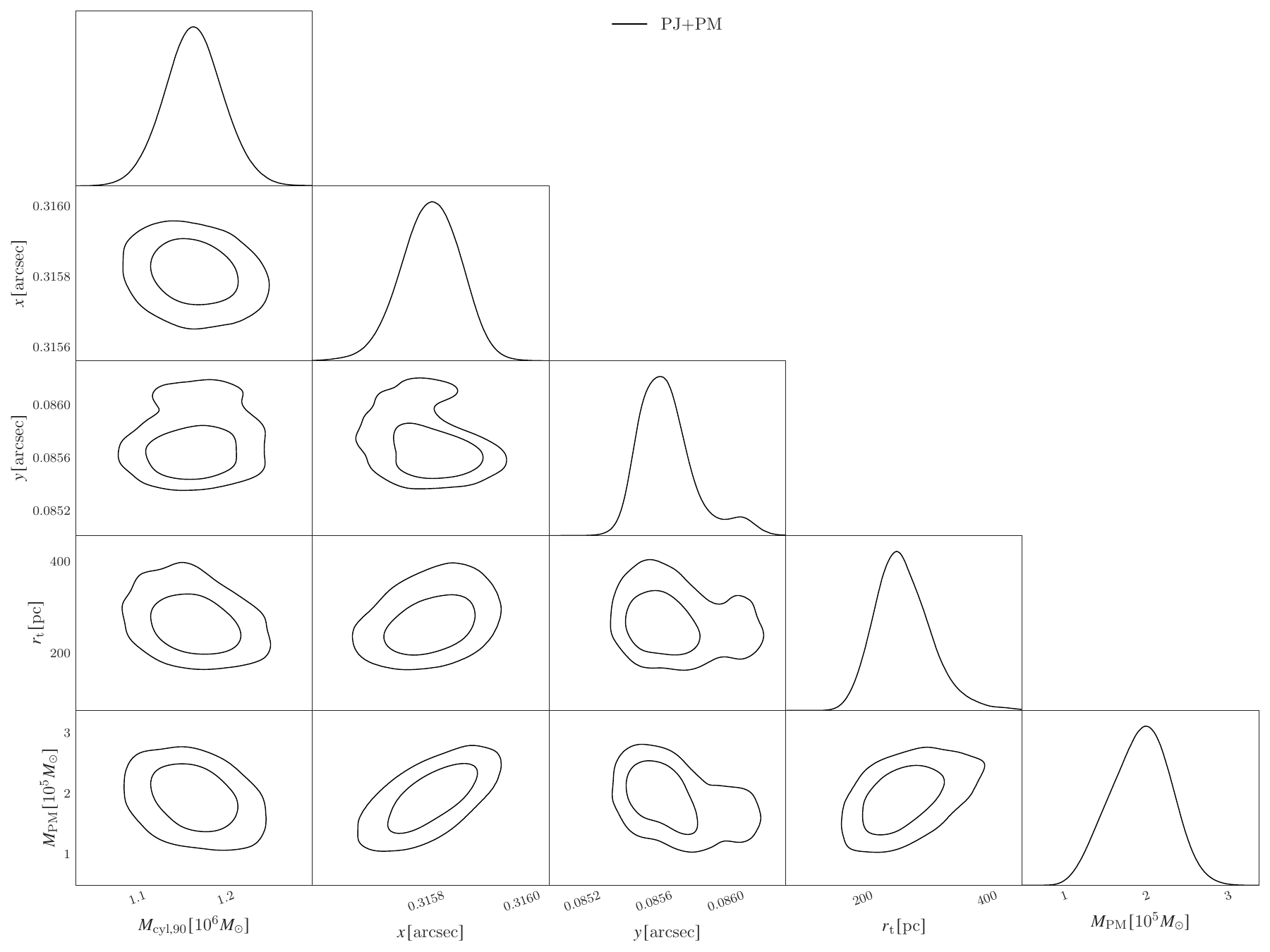}
\caption{{\bf Posterior distributions for the parameters of the Pseudo-Jaffe plus point-mass model.} The contours represent the 1- and 2-$\sigma$ confidence regions.} 
\end{figure}

\begin{figure}[h]
\includegraphics[width=1.4\textwidth, angle=90]{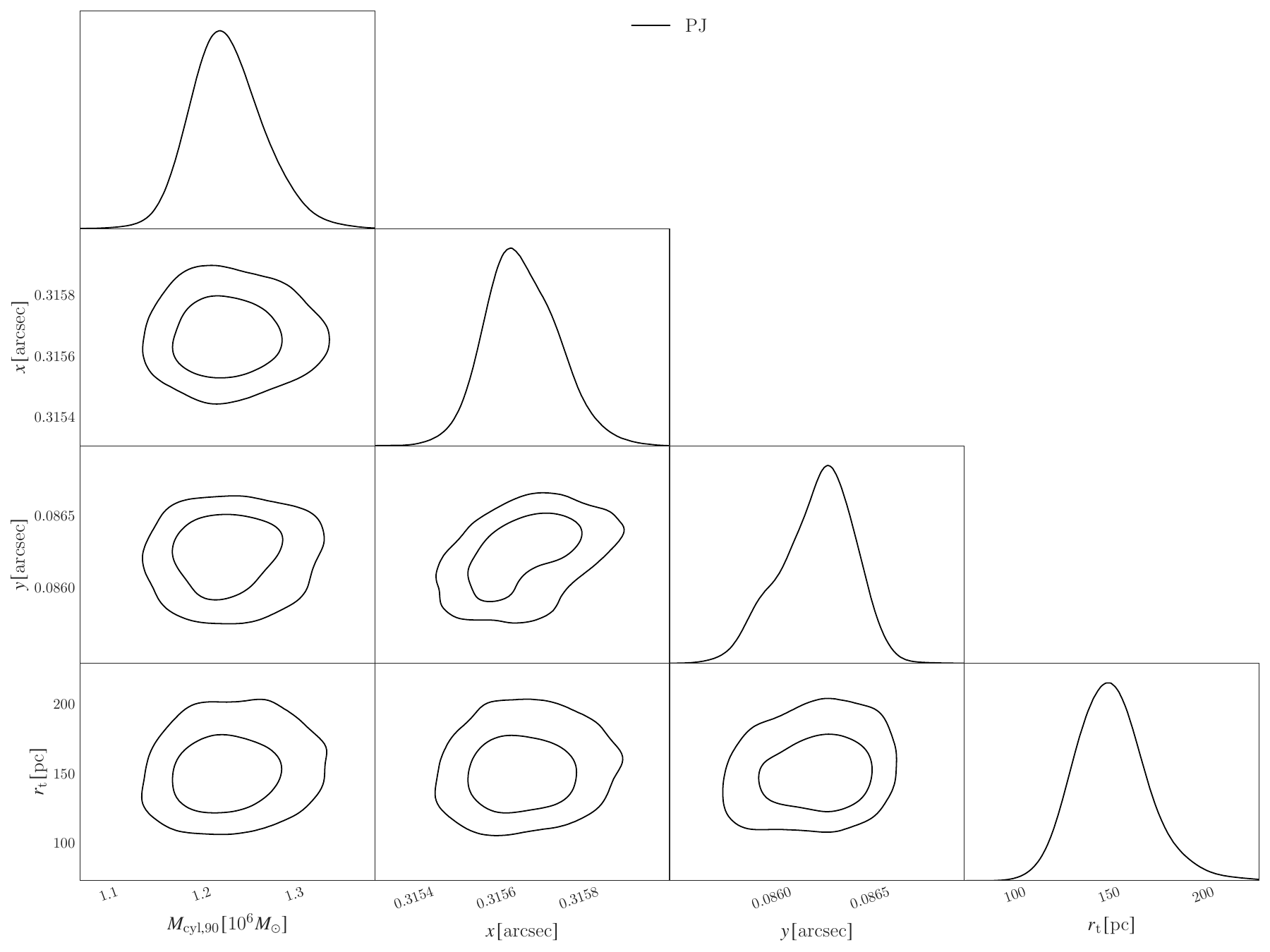}
\caption{{\bf Posterior distributions for the parameters of the Pseudo-Jaffe model.} The contours represent the 1- and 2-$\sigma$ confidence regions.} 
\end{figure}

\begin{figure}[h]
\includegraphics[width=1.4\textwidth, angle=90]{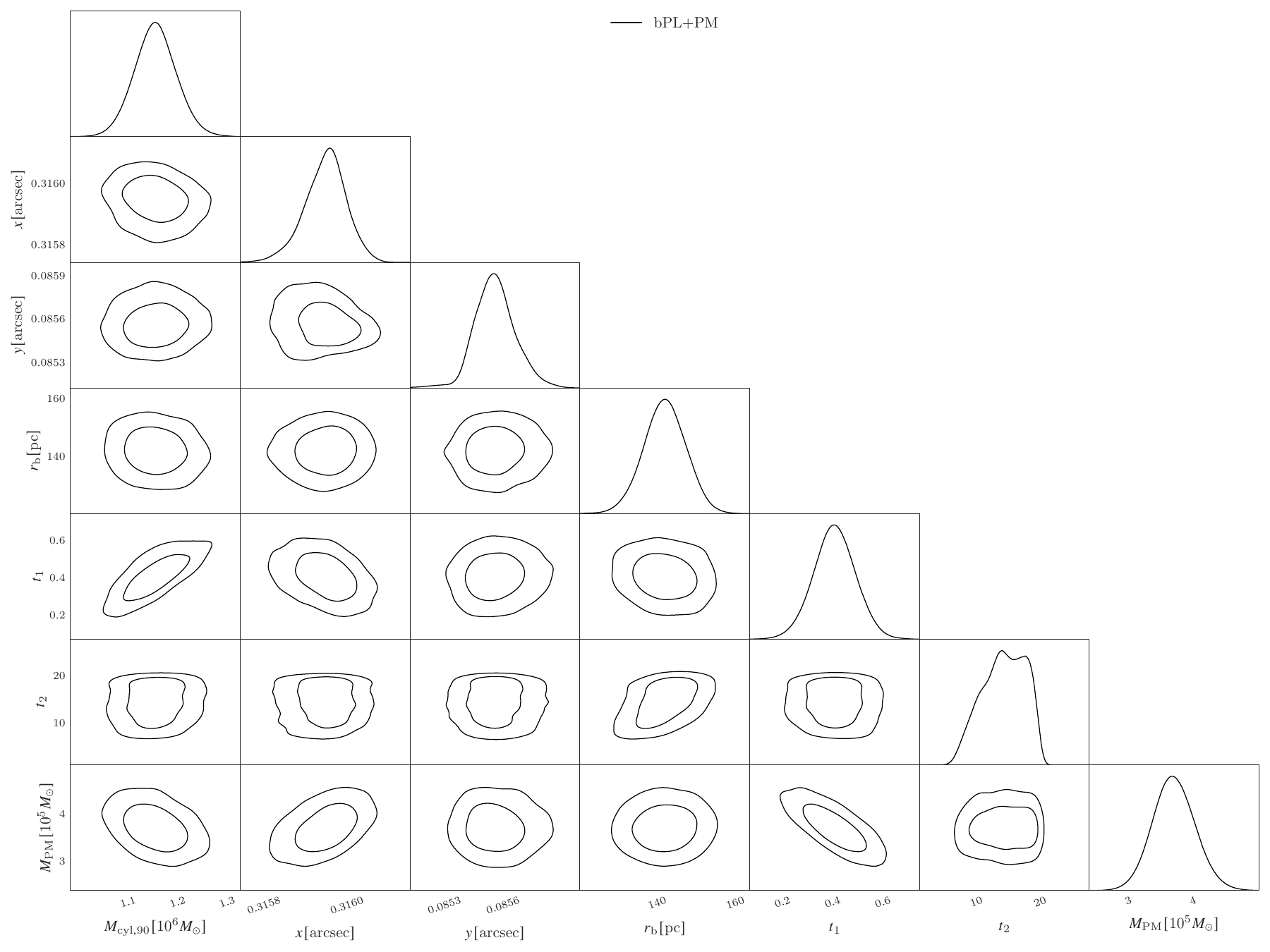}
\caption{{\bf Posterior distributions for the parameters of the broken power-law plus point-mass model.} The contours represent the 1- and 2-$\sigma$ confidence regions.} 
\end{figure}

\begin{figure}[h]
\includegraphics[width=1.4\textwidth, angle=90]{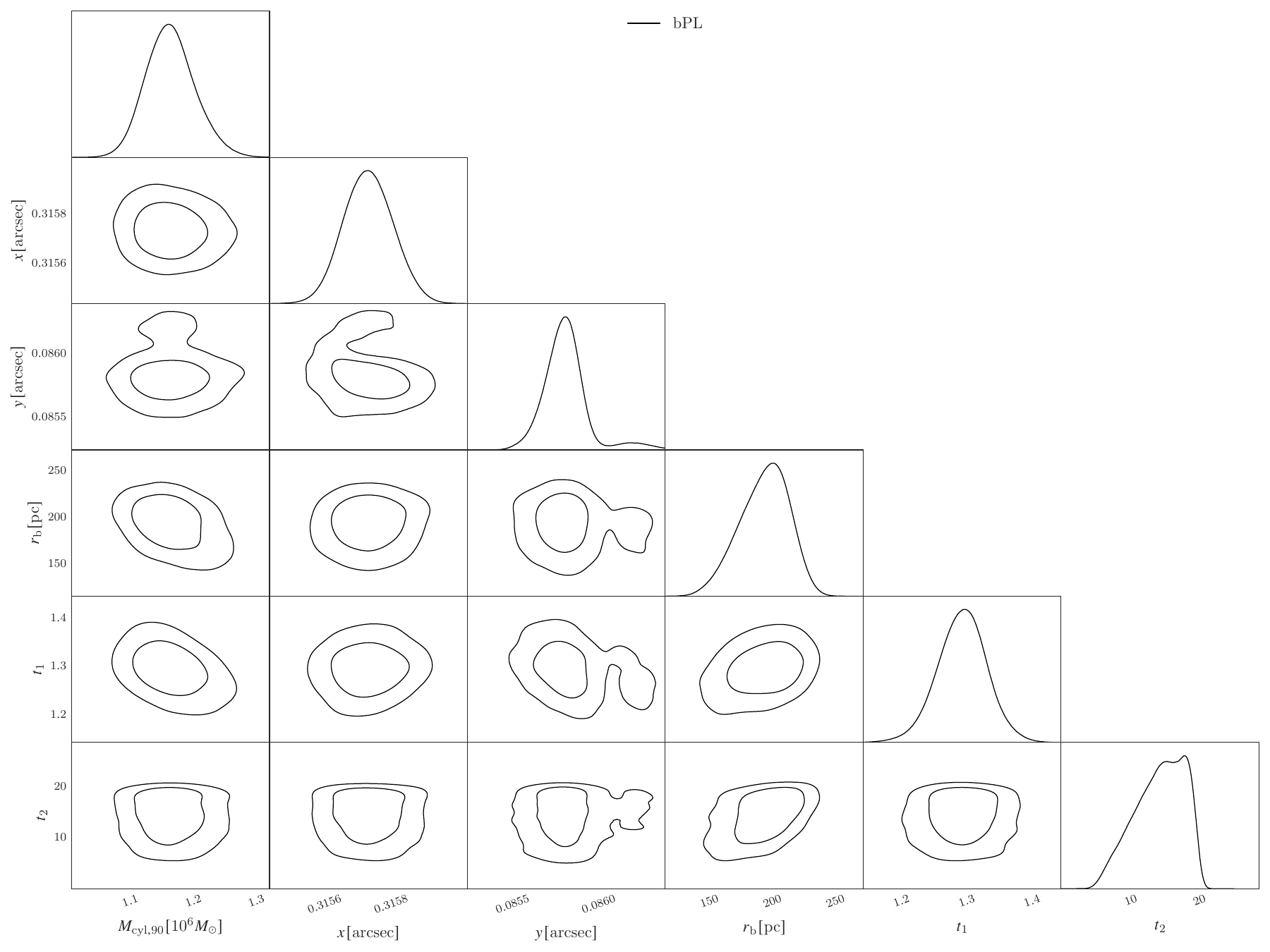}
\caption{{\bf Posterior distributions for the parameters of the broken power-law model.} The contours represent the 1- and 2-$\sigma$ confidence regions.} 
\end{figure}

\begin{figure}[h]
\includegraphics[width=1.4\textwidth, angle=90]{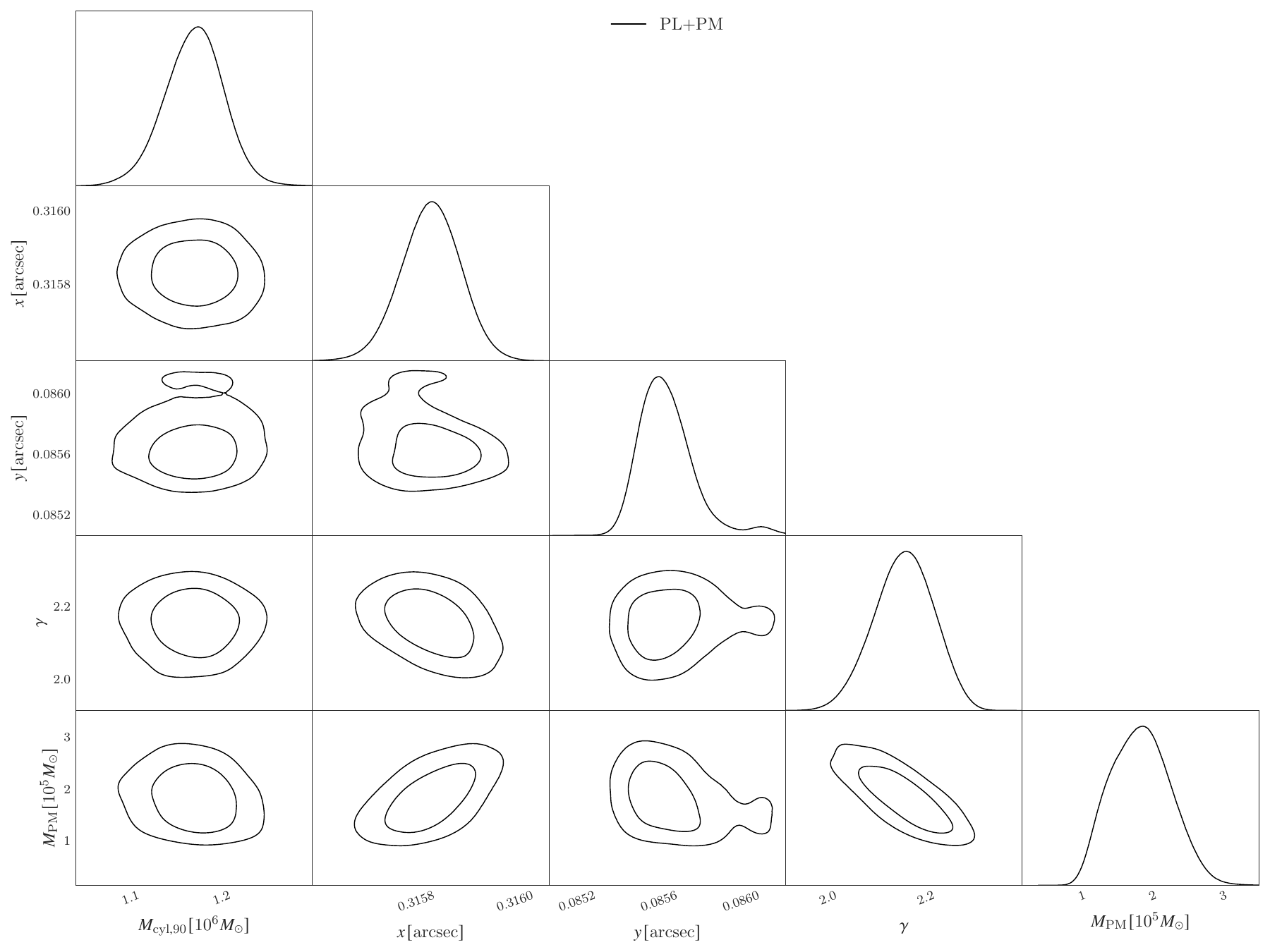}
\caption{{\bf Posterior distributions for the parameters of the power-law plus point-mass model.} The contours represent the 1- and 2-$\sigma$ confidence regions.} 
\end{figure}

\begin{figure}[h]
\includegraphics[width=1.4\textwidth, angle=90]{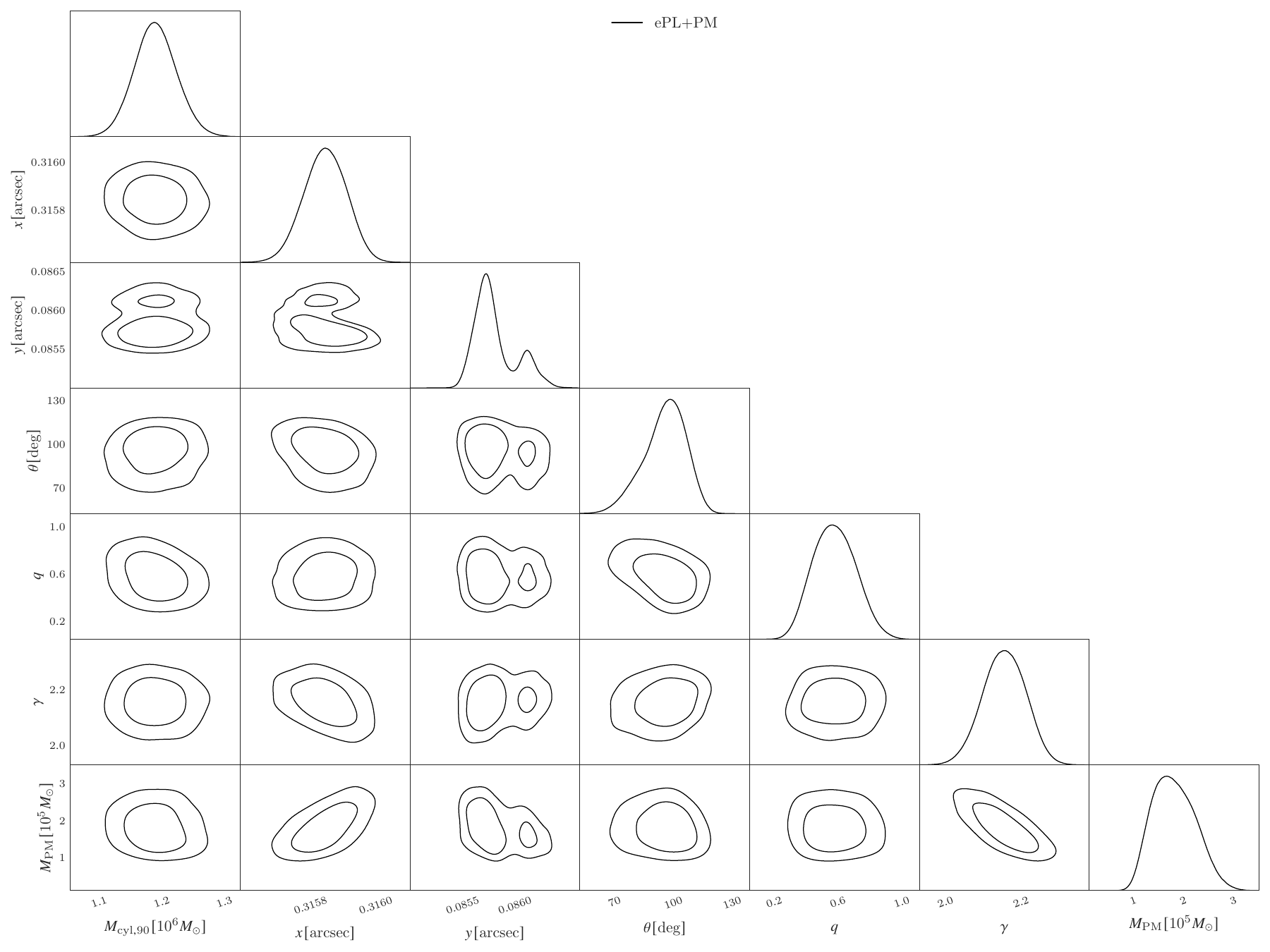}
\caption{{\bf Posterior distributions for the parameters of the elliptical power-law plus point-mass model.} The contours represent the 1- and 2-$\sigma$ confidence regions.} 
\end{figure}

\begin{figure}[h]
\includegraphics[width=1.4\textwidth, angle=90]{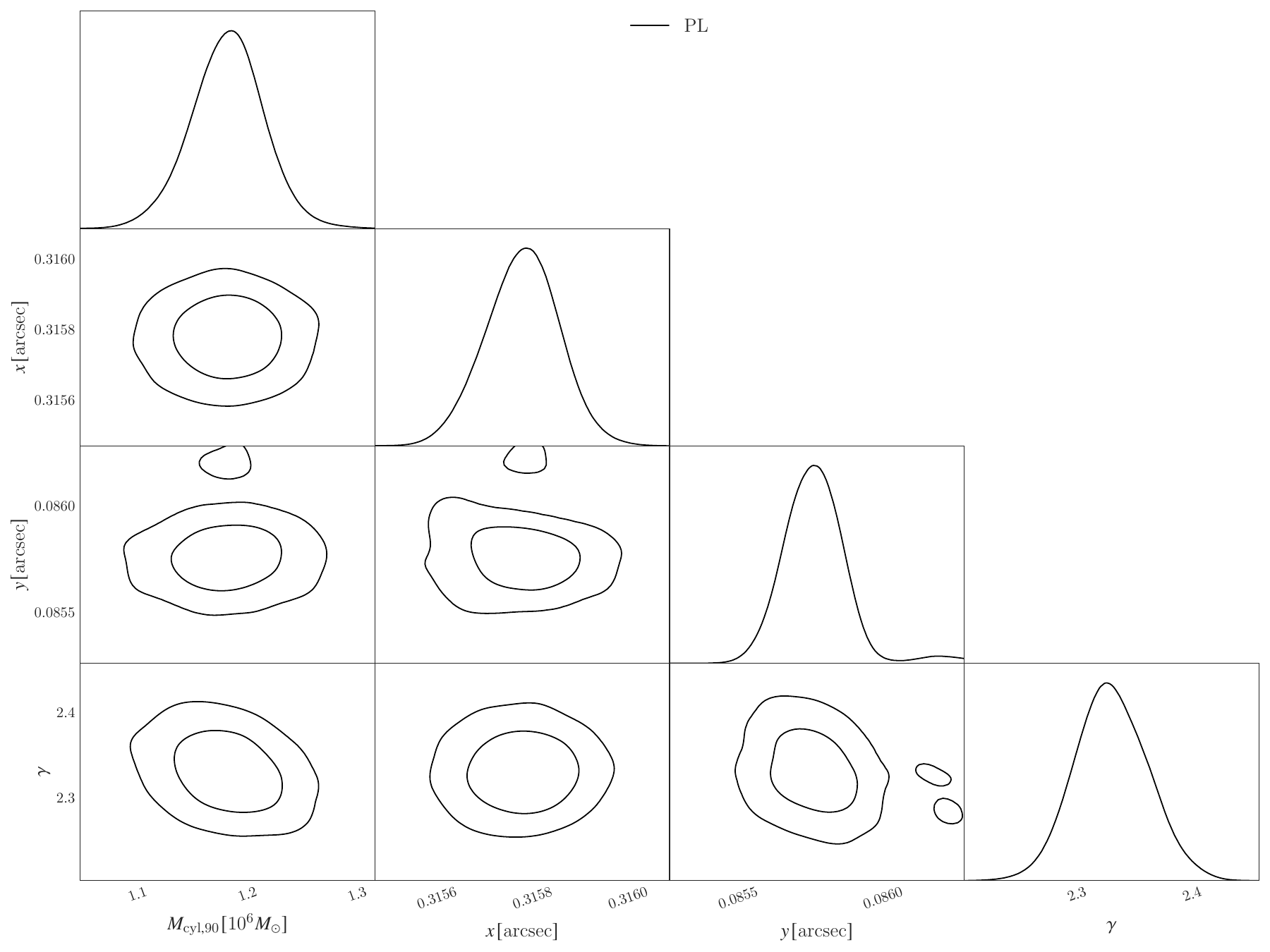}
\caption{{\bf Posterior distributions for the parameters of the power-law model.} The contours represent the 1- and 2-$\sigma$ confidence regions.} 
\end{figure}

\begin{figure}[h]
\includegraphics[width=1.4\textwidth, angle=90]{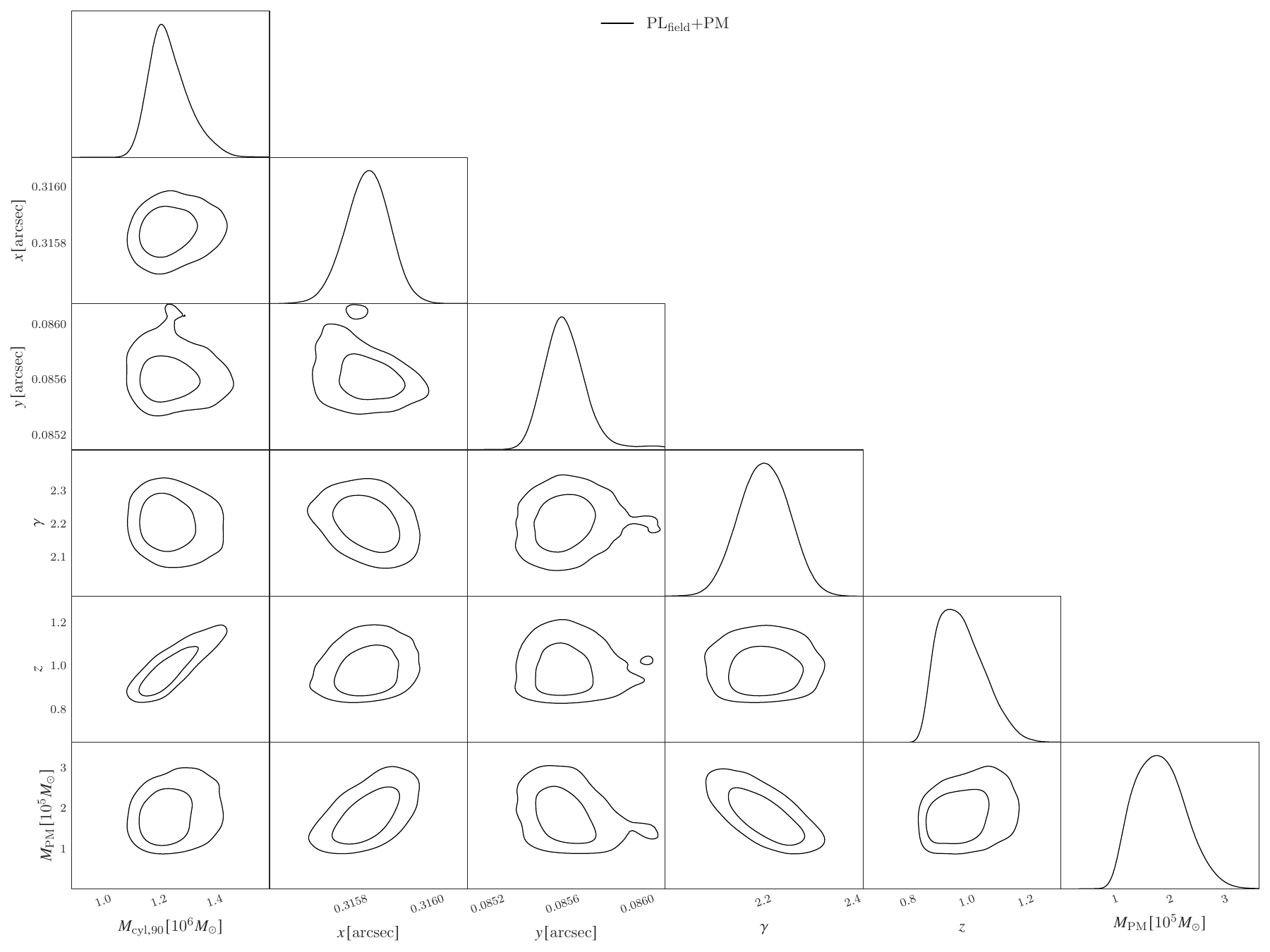}
\caption{{\bf Posterior distributions for the parameters of the free-redshift power-law plus point mass model.} The contours represent the 1- and 2-$\sigma$ confidence regions.} 
\end{figure}

\begin{figure}[h]
\includegraphics[width=1.4\textwidth, angle=90]{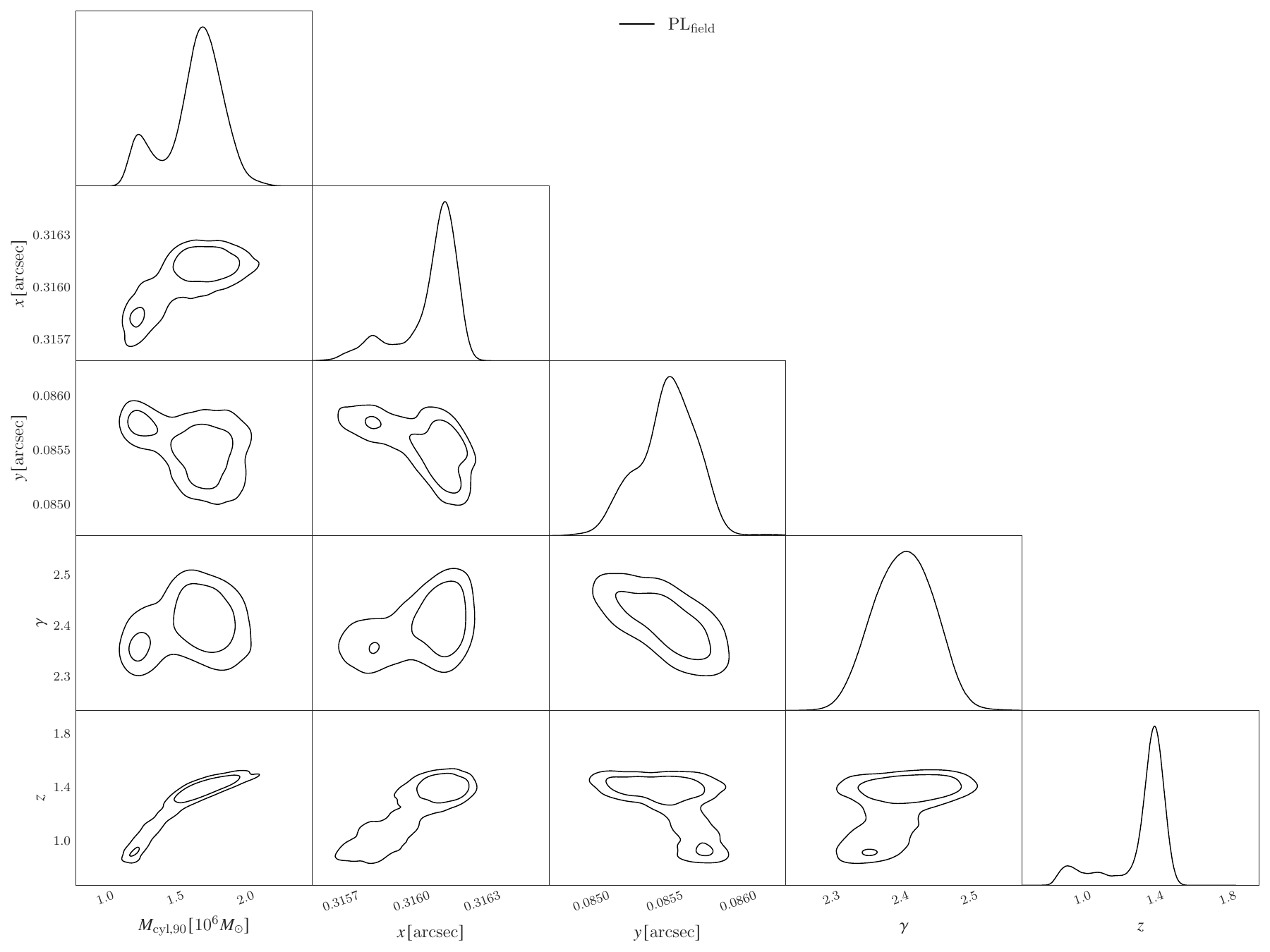}
\caption{{\bf Posterior distributions for the parameters of the free-redshift power-law model.} The contours represent the 1- and 2-$\sigma$ confidence regions.} 
\end{figure}

\begin{figure}[h]
\includegraphics[width=1.4\textwidth, angle=90]{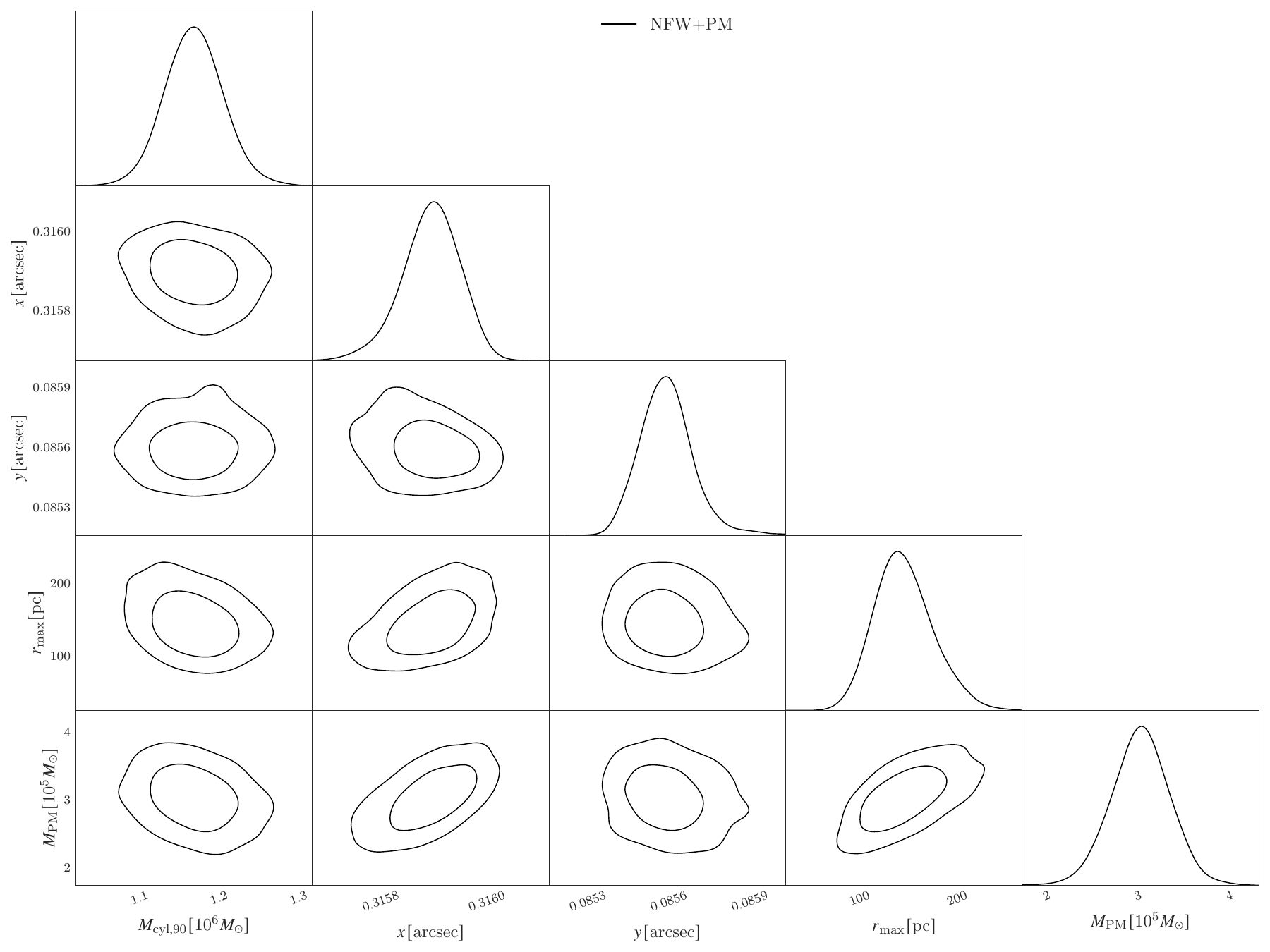}
\caption{{\bf Posterior distributions for the parameters of the NFW plus point mass model.} The contours represent the 1- and 2-$\sigma$ confidence regions.} 
\end{figure}

\begin{figure}[h]
\includegraphics[width=1.4\textwidth, angle=90]{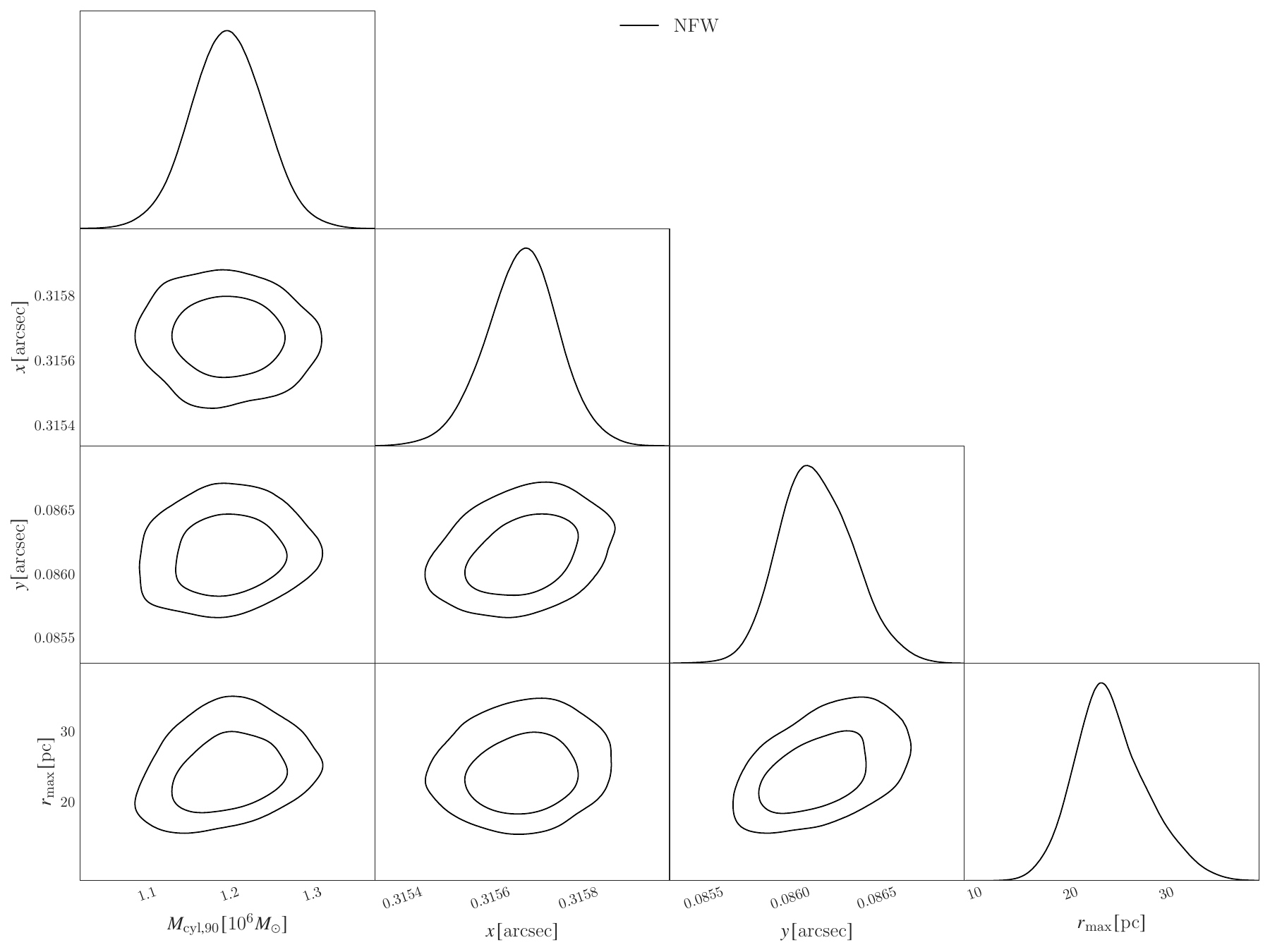}
\caption{{\bf Posterior distributions for the parameters of the NFW model.} The contours represent the 1- and 2-$\sigma$ confidence regions.} 
\end{figure}

\begin{figure}[h]
\includegraphics[width=1.4\textwidth, angle=90]{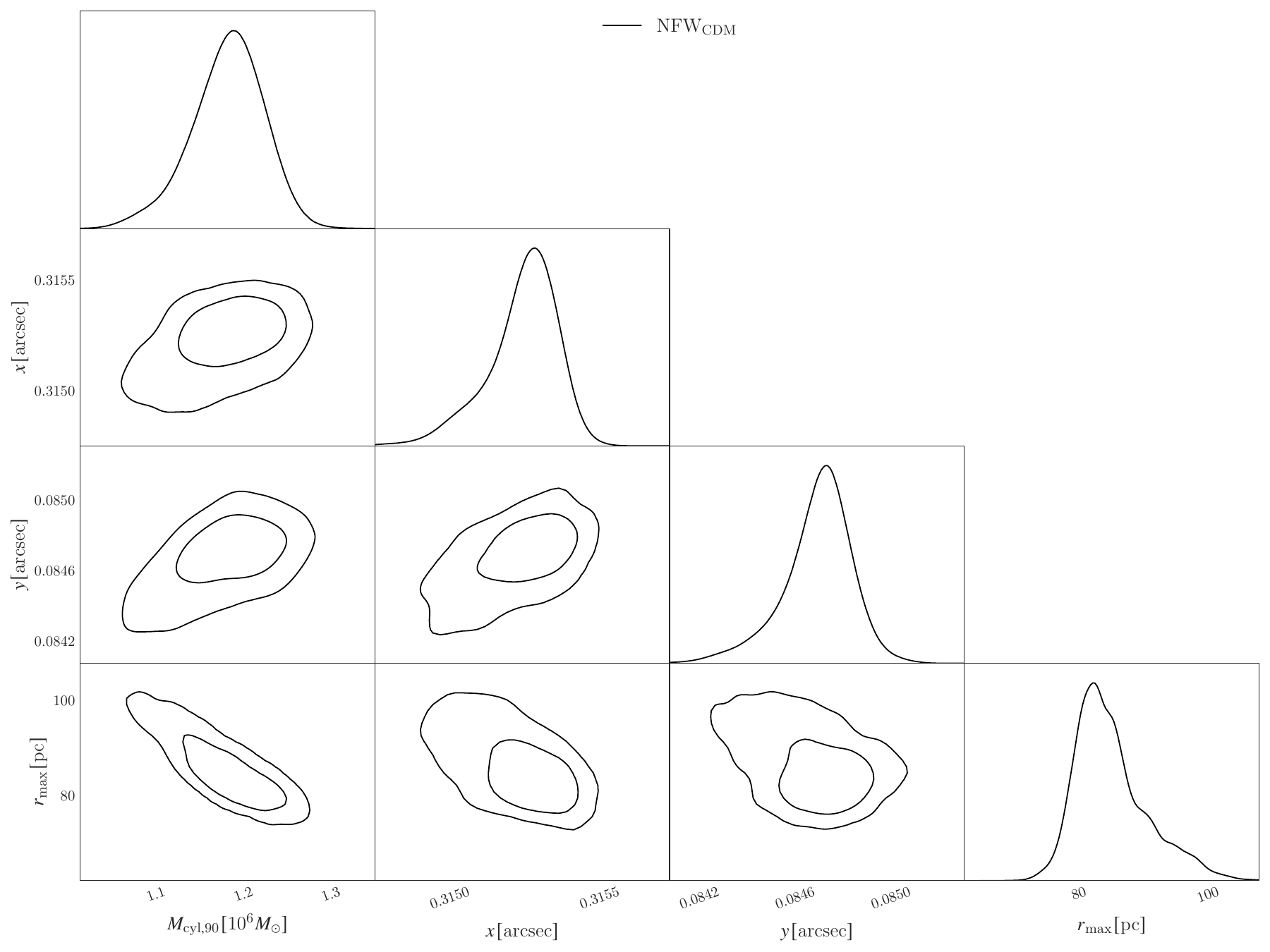}
\caption{{\bf Posterior distributions for the parameters of the NFW$_{\rm CDM}$ model.} The contours represent the 1- and 2-$\sigma$ confidence regions. The prior on r$_{\rm max}$ is informed by CDM simulations.} 
\end{figure}

\begin{figure}[h]
\includegraphics[width=1.4\textwidth, angle=90]{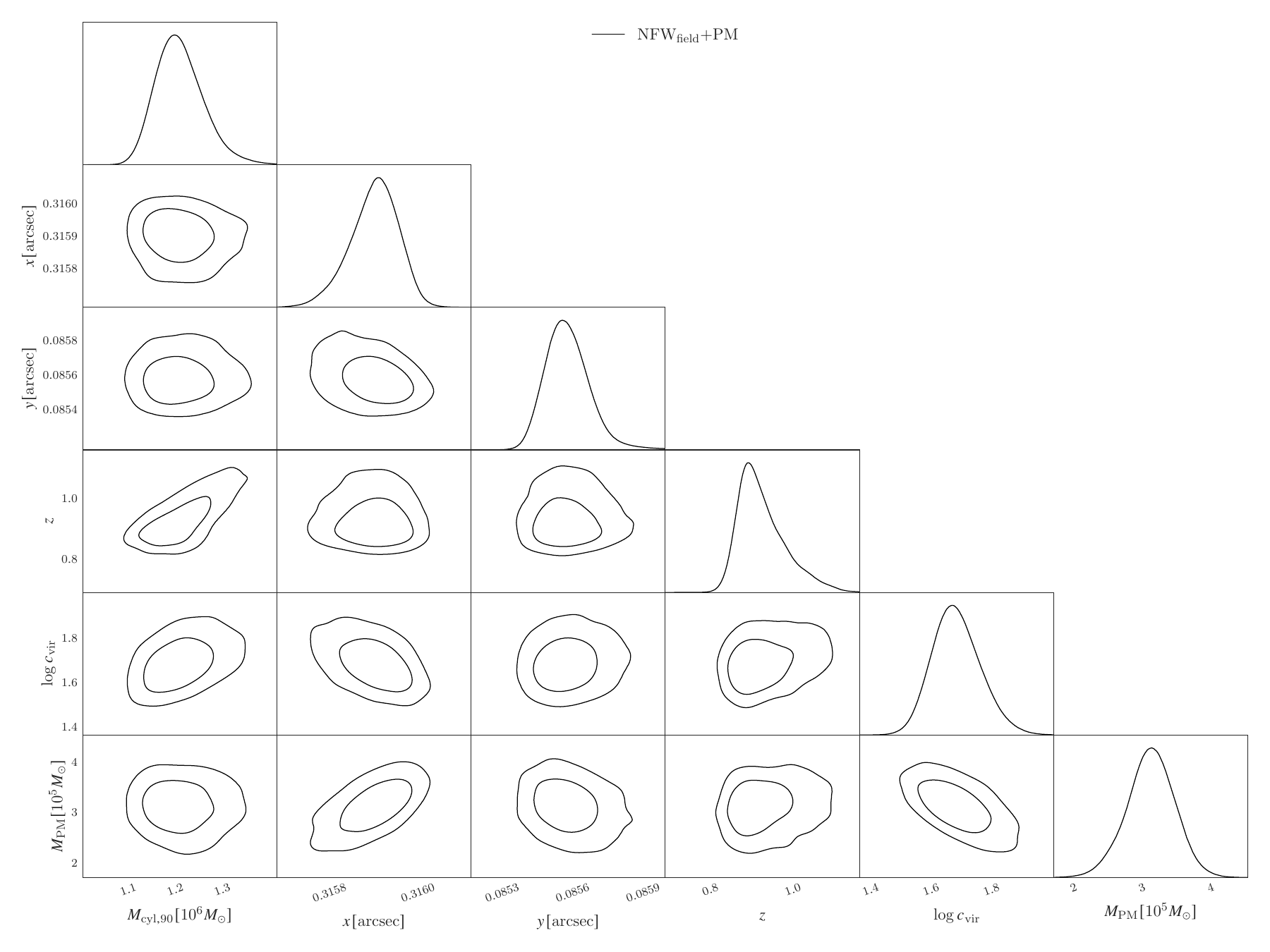}
\caption{{\bf Posterior distributions for the parameters of the free-redshift NFW plus point mass model.} The contours represent the 1- and 2-$\sigma$ confidence regions.} 
\end{figure}

\begin{figure}[h]
\includegraphics[width=1.4\textwidth, angle=90]{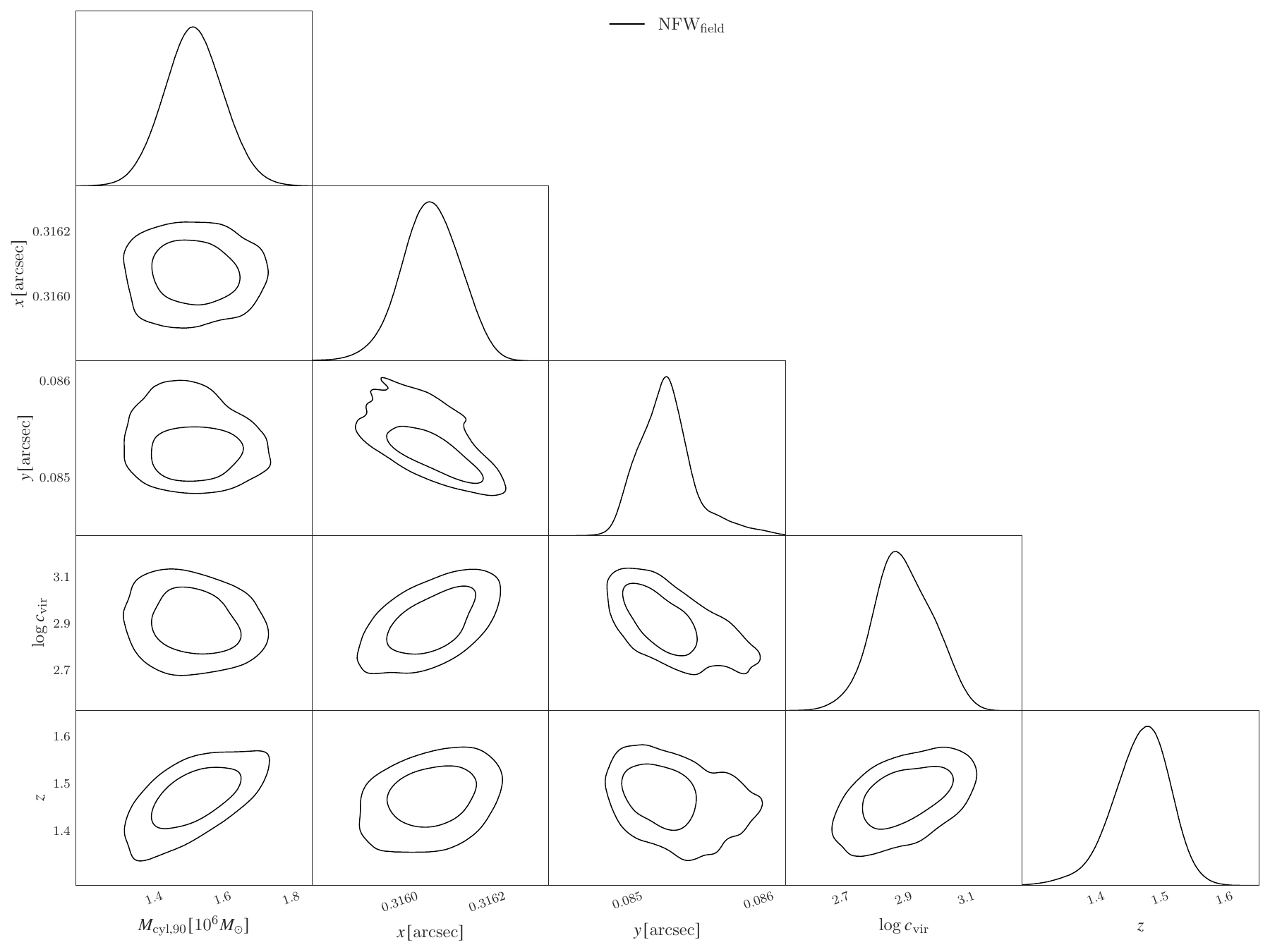}
\caption{{\bf Posterior distributions for the parameters of the free-redshift NFW model.} The contours represent the 1- and 2-$\sigma$ confidence regions.} 
\end{figure}

\begin{figure}[h]
\includegraphics[width=1.4\textwidth, angle=90]{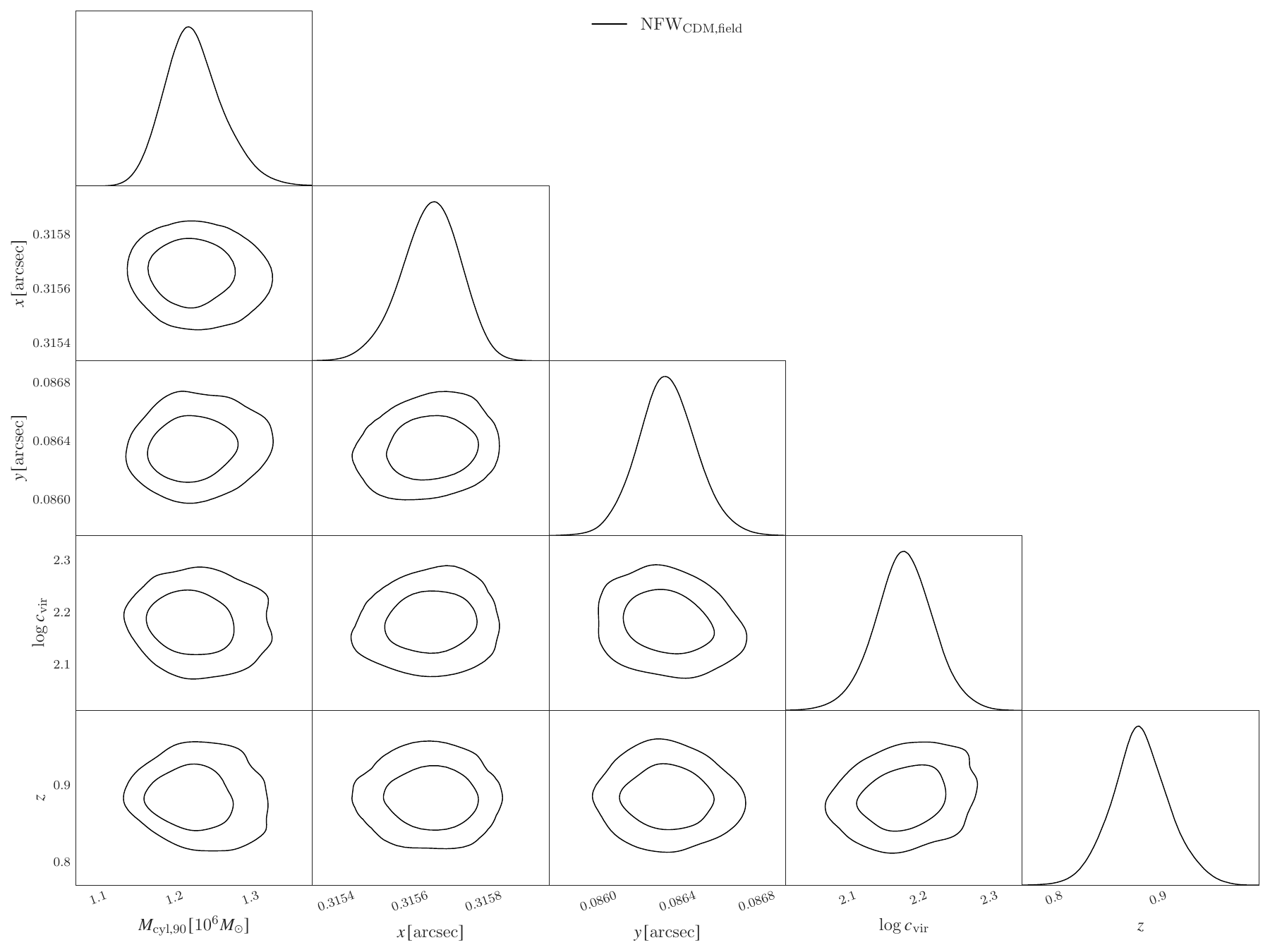}
\caption{{\bf Posterior distributions for the parameters of the free-redshift NFW$_{\rm CDM}$ model.} The contours represent the 1- and 2-$\sigma$ confidence regions.  The prior on the concentration is informed by CDM simulations.} 
\end{figure}

\begin{figure}[h]
\includegraphics[width=1.4\textwidth, angle=90]{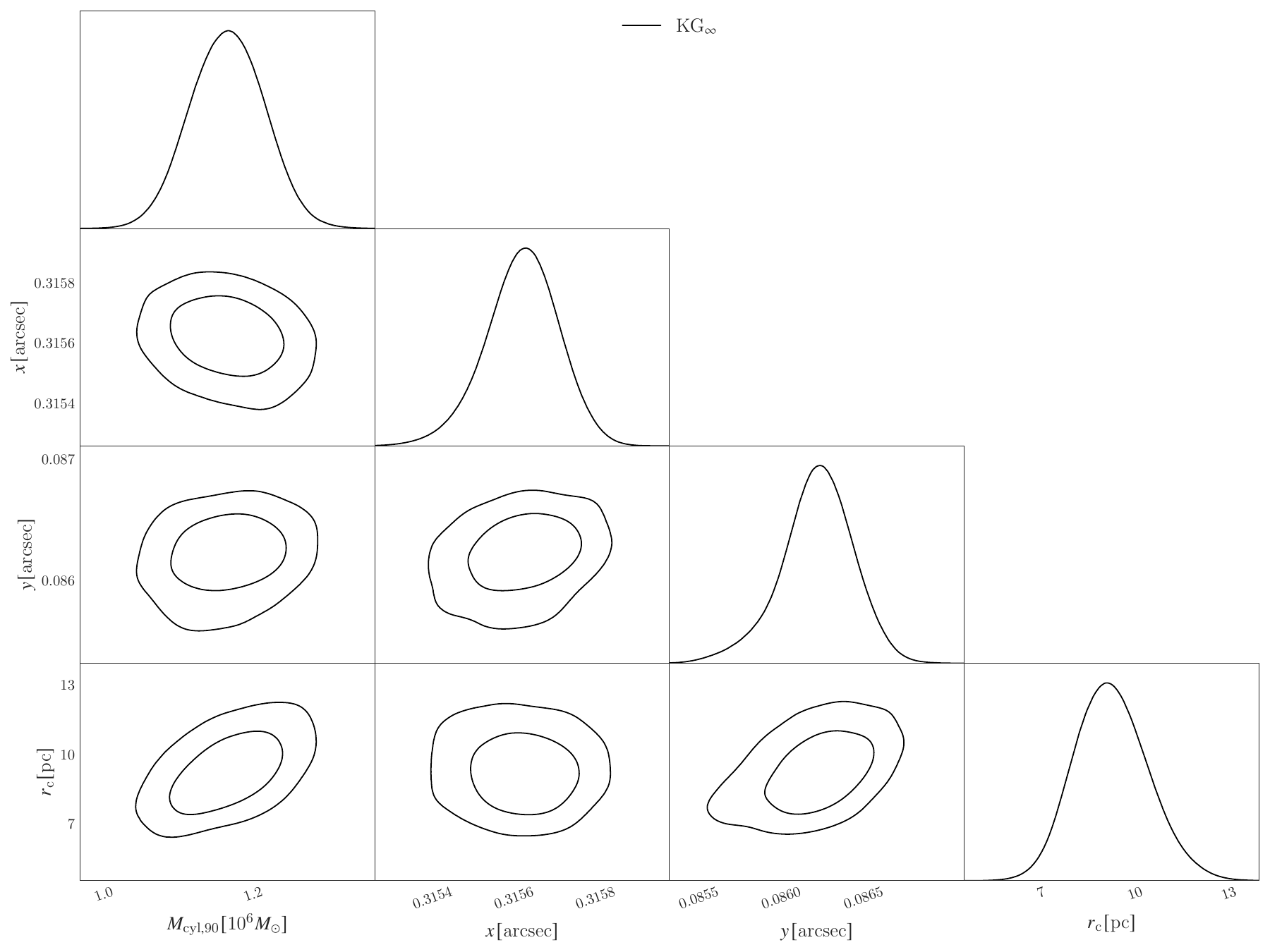}
\caption{{\bf Posterior distributions for the parameters of the KG$_{\infty}$ model.} The contours represent the 1- and 2-$\sigma$ confidence regions. The truncation radius is assumed to $r_{\rm t}\rightarrow\infty$.} 
\end{figure}

\begin{figure}[h]
\includegraphics[width=1.4\textwidth, angle=90]{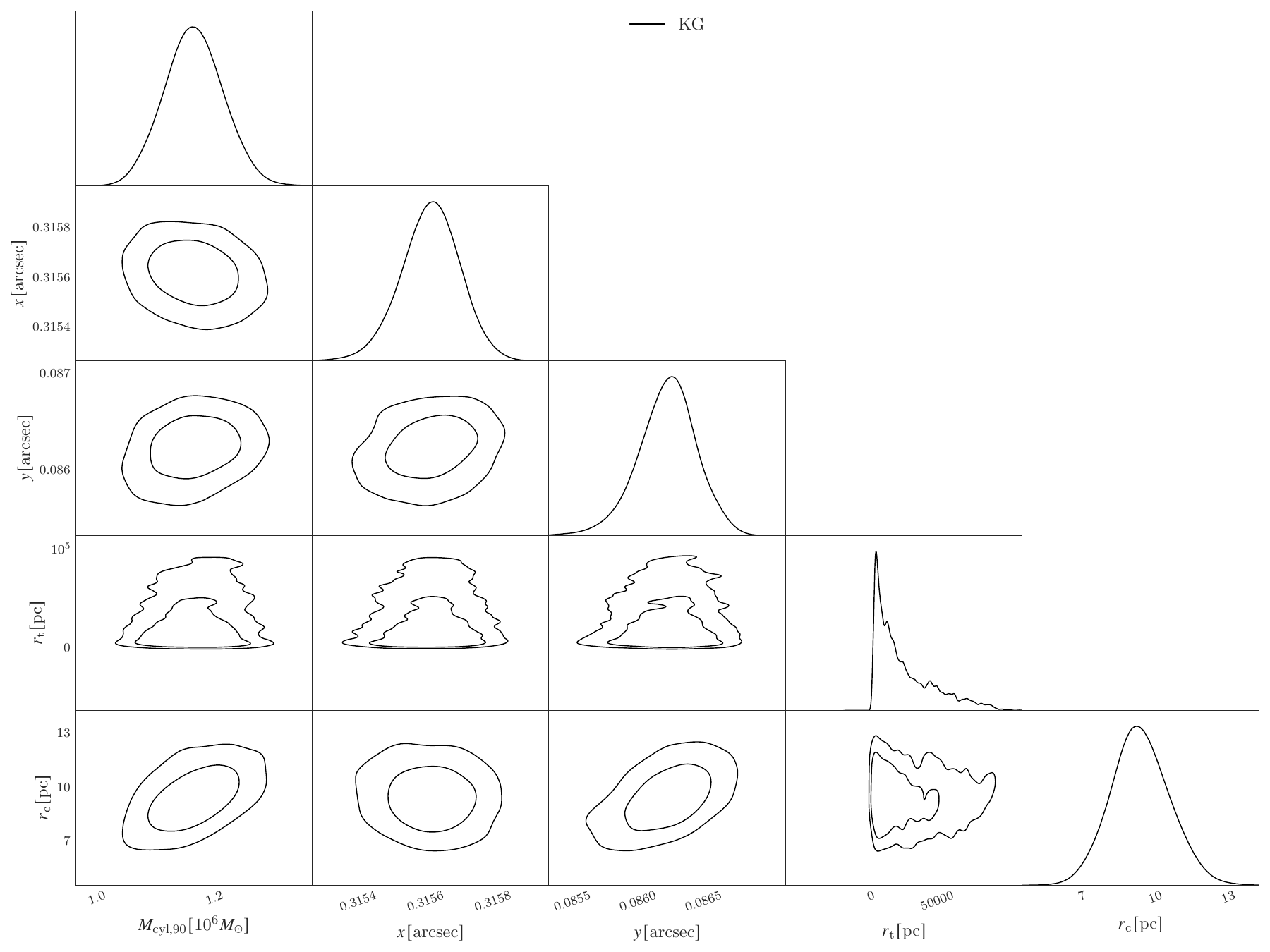}
\caption{{\bf Posterior distributions for the parameters of the KG model.} The contours represent the 1- and 2-$\sigma$ confidence regions.} 
\end{figure}

\begin{figure}[h]
\includegraphics[width=1.1\textwidth, angle=90]{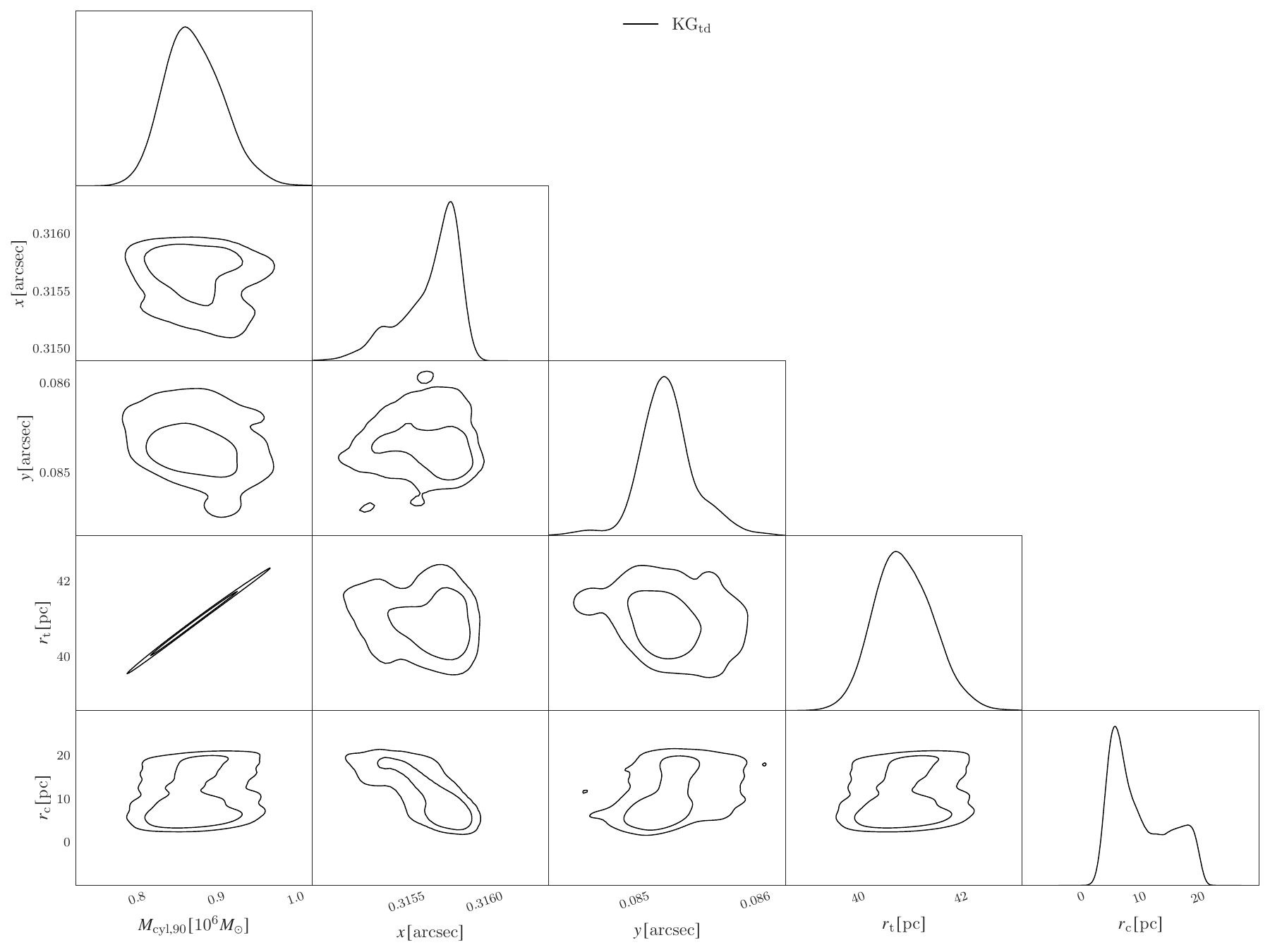}
\caption{{\bf Posterior distributions for the parameters of the KG$_{\rm td}$ model.} The contours represent the 1- and 2-$\sigma$ confidence regions. The truncation radius is assumed to be the same as the tidal radius.} 
\end{figure}

\begin{figure}[h]
\includegraphics[width=1.4\textwidth, angle=90]{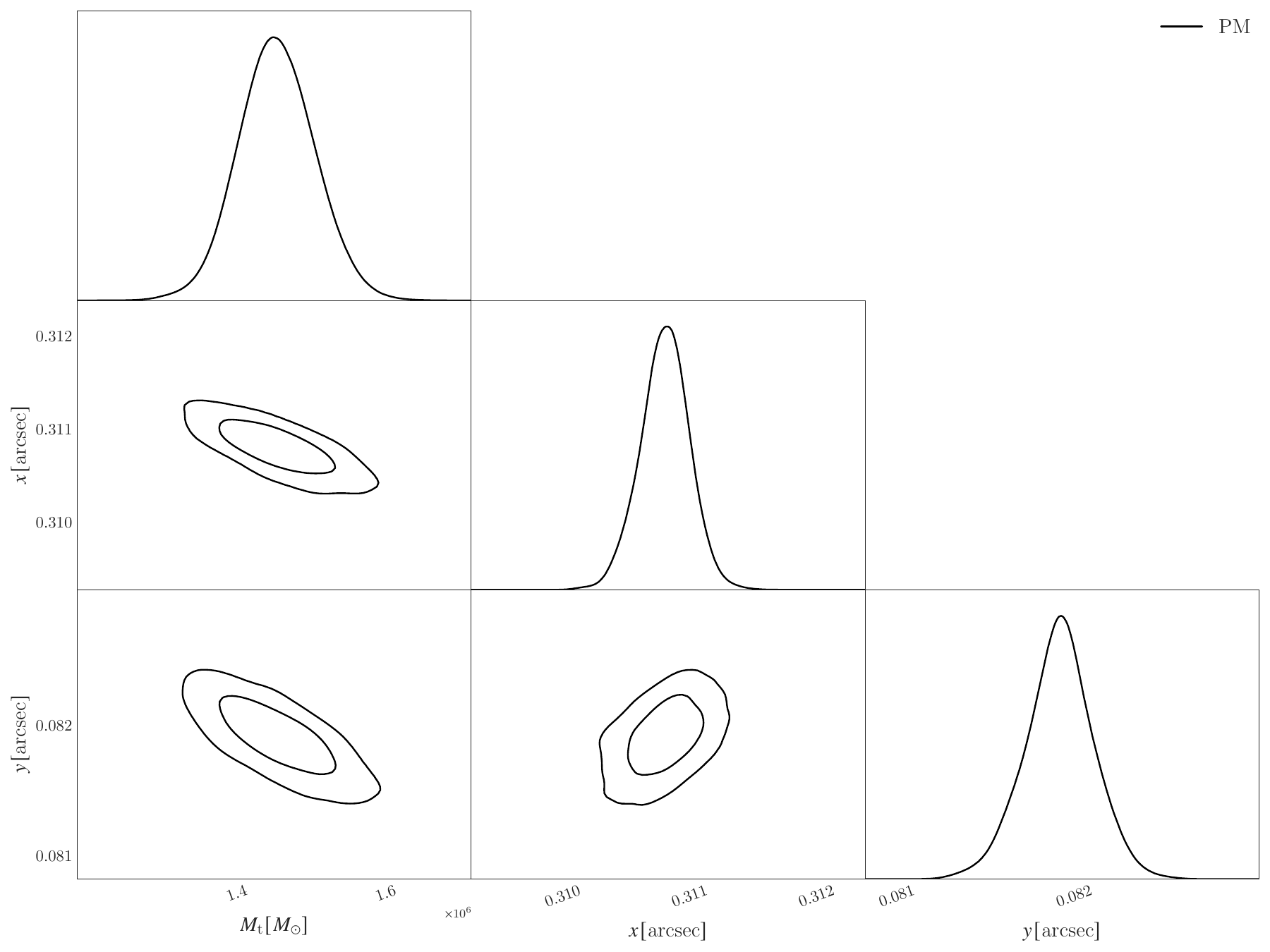}
\caption{{\bf Posterior distributions for the parameters of the point mass model.} The contours represent the 1- and 2-$\sigma$ confidence regions.} 
\label{fig:si_fig24}
\end{figure}

\end{document}